\documentclass[1p]{elsarticle}

\usepackage{lipsum}
\makeatletter
\def\ps@pprintTitle{%
 \let\@oddhead\@empty
 \let\@evenhead\@empty
 \def\@oddfoot{}%
 \let\@evenfoot\@oddfoot}
\makeatother

\usepackage{amssymb}
\usepackage{natbib}
\usepackage{bm}
\usepackage{amsmath}

\usepackage{subfigure}

\usepackage{multirow}

\usepackage{color}
\usepackage{siunitx}

\usepackage[boxed,ruled,linesnumbered]{algorithm2e}

\usepackage{mathrsfs}
\usepackage{color}
\usepackage[normalem]{ulem}

\journal{Journal of Computational Physics}

\begin{document}

\begin{frontmatter}

\title{Minimal subspace rotation on the Stiefel manifold for stabilization and
enhancement of projection-based reduced order models for the compressible
Navier-Stokes equations}

\author[aa]{Maciej
Balajewicz\corref{cor1}\fnref{fn1}}\ead{mbalajew@illinois.edu}

\author[bb]{Irina Tezaur\fnref{fn2}}
\author[cc]{Earl Dowell\fnref{fn3}}

\address[aa]{Department of Aerospace Engineering,
         University of Illinois at Urbana-Champaign, Urbana, IL, USA}
\address[bb]{Quantitative Modeling \& Analysis
Department, Sandia National Laboratories, P.O. Box 969, MS 9159, Livermore, CA, USA}
\address[cc]{Department of Mechanical Engineering and Materials
Science, Duke University, Durham, NC, USA}

\fntext[fn1]{Assistant Professor}
\fntext[fn2]{Principal Member of Technical Staff}
\fntext[fn3]{William Holland Hall Professor}
\cortext[cor1]{Corresponding author}

\begin{abstract}
For a projection-based reduced order model (ROM) of a fluid flow to be stable and
accurate, the dynamics of the truncated subspace must be taken into account.
This paper proposes an approach for stabilizing and
enhancing projection-based fluid ROMs in which truncated modes are
accounted for \textit{a priori} via a minimal rotation of the
projection subspace.  Attention is focused on the full non-linear
compressible Navier-Stokes equations in specific volume form
as a step toward a more general formulation for problems with generic
non-linearities.  Unlike traditional approaches, no empirical
turbulence modeling terms are required, and consistency between the
ROM and the full order model from which the ROM is derived is
maintained.  Mathematically, the approach is formulated as a
trace minimization problem on the Stiefel manifold. The reproductive
as well as predictive capabilities of the method are evaluated on
several compressible flow problems, including a problem involving
laminar flow over an airfoil with a high angle of attack, and a
channel-driven cavity flow problem.
\end{abstract}

\end{frontmatter}

\section{Introduction}
\label{sec:Introduction}

The past several decades have seen an exponential growth of computer processing
speed and memory capacity.  The massive, complex simulations that run on
supercomputers allow exploration of fields for which physical experiments are
too impractical, hazardous, and/or costly. A striking number of these fields
require computational fluid dynamics (CFD) models and simulations. Accurate and
efficient CFD simulations are critical to many defense, climate and energy
missions, e.g., simulations aimed to design and qualify nuclear weapons
components carried within an aircraft weapons bay; global climate simulations
aimed to predict anticipated twenty-first century sea-level rise; aero-elastic
simulations for optimal design of wind systems for power generation.

Unfortunately, even with the aid of massively parallel
next-generation computers, CFD simulations for applications such as
these are still too expensive for real-time and multi-query
applications such as uncertainty quantification (UQ), optimization
and control design. Reduced order modeling is a promising tool for
bridging the gap between high-fidelity, and real-time
simulations/UQ.  Reduced order models (ROMs) are derived from a
sequence of high-fidelity simulations but have a much lower
computational cost.  Hence, ROMs can enable real-time simulations of
complex systems for more rapid analysis, control and
decision-making in the presence of uncertainty.

Most existing ROM approaches are based on projection.  In
projection-based reduced order modeling, the state variables are
approximated in a low-dimensional subspace.  There exist a number of
approaches for calculating this low-dimensional subspace, e.g.,
Proper Orthogonal Decomposition (POD)~\cite{Sirovich_QAP_1987,
Holmes_BOOK_2012}, Dynamic Mode Decomposition
(DMD)~\citep{Rowley_JFM_2009,Schmid_JFM_2010} balanced POD
(BPOD)~\cite{Rowley_IJBC_2005, Willcox_AIAA_2002}, balanced
truncation~\cite{Gugercin_IJC_2004, Moore_IEEETAC_1981}, and the
reduced basis method \cite{Rozza_CCP_2011, Veroy_IJNMF_2005}. In all
of these methods, a basis for the low-dimensional subspace is
obtained from a basis for a higher-dimensional subspace through
truncation -- the removal of modes that are believed to be
unimportant in representing a problem solution.
Typically, the size of the reduced basis is chosen according to an
energy criterion: modes with low energy are discarded, so that the
reduced basis subspace is spanned by the highest energy modes.
Although truncated modes are
negligible from a data compression point of view, they are often
crucial for representing solutions to the dynamical flow equations.
Dynamics of the truncated orthogonal subspace must be taken into account for
to ensure stability and accuracy of the ROM.

For linear systems, a variety of techniques for generating low-dimensional
projection- based ROMs with rigorous stability guarantees and accuracy bounds
are available~\cite{Gugercin_IJC_2004, Moore_IEEETAC_1981,Amsallem_IJNME_2012,
Kalashnikova_CMAME_2014}.  Equivalent results are lacking for nonlinear systems.
Traditionally, low-dimensional ROMs of fluid flows have been stabilized and
enhanced using empirical turbulence models. In this approach, the nonlinear
dynamics of the truncated subspace are modeled using additional constant and
linear terms in the ROM system of ordinary differential equations
(ODEs)~\cite{Aubry_JFM_1988,Rempfer_JFM_1994,Cazemier_PF_1998,Galletti_JFM_2004}.
More recently, nonlinear eddy-viscosity models have also been
proposed~\cite{Osth_JFM_2014,Noack_JNET_2008,Iliescu_NMPDE_2014,Protas_ARXIV_2014}.
One downside of turbulence models is that they destroy consistency between the
Navier-Stokes partial differential equations (PDEs) and the ODE system of the
ROM. Accurately identifying and matching free coefficients of the turbulence
models is another challenge.  Moreover, these methods are usually limited to the
incompressible Navier-Stokes equations.

Consider, for concreteness, the POD/Galerkin approach to model reduction applied
to the incompressible Navier-Stokes equations.  For these equations, the natural
choice of inner product for the Galerkin projection step of the model reduction
procedure is the $L^2$ inner product. This is because, in these models, the
solution vector is taken to be the velocity vector $\mathbf{u}$, so that
$||\mathbf{u}||_2$ is a measure of the global kinetic energy in the domain, and
the POD modes optimally represent the kinetic energy present in the ensemble
from which they were generated. The same is not true for the compressible
Navier-Stokes equations. Hence, if a compressible fluid ROM is constructed in
the $L^2$ inner product (a common choice of inner product in projection-based
model reduction), the ROM solution may not satisfy the conservation relation
implied by the governing equations, and may exhibit non-physical
instabilities~\cite{Rowley_PDNP_2004}.

Unfortunately, ROM instability is a real problem for many compressible flow
problems: as demonstrated in~\cite{Barone_SAND_2015, Bui_JCP_2007,
Barone_JCP_2009, Kalashnikova_SAND_2014}, a compressible fluid POD/Galerkin ROM
might be stable for a given number of modes, but unstable for other choices of
basis size. Several researchers have proposed ways to circumvent this difficulty
through the careful construction of an energy-based inner product for the
projection step of the model reduction. \citet{Rowley_PDNP_2004}  show that
Galerkin projection preserves the stability of an equilibrium point at the
origin if the ROM is constructed in an energy--based inner product.
\citet{Barone_JCP_2009, Kalashnikova_SAND_2014} demonstrate that a symmetry
transformation leads to a stable formulation for a Galerkin ROM for the
linearized compressible Euler equations and nonlinear compressible Navier-Stokes
equations with solid wall and far-field boundary conditions.
\citet{Serre_JCP_2012} propose applying the stabilizing projection developed
by~\citet{Barone_JCP_2009,Kalashnikova_SAND_2014} to a skew-symmetric system
constructed by augmenting a given linear system with its adjoint system. The
downside to these methods is that they are inherently embedded methods: access
to the governing PDEs and/or the code that discretizes these PDEs is required.

Other ROM approaches, e.g., the Gauss-Newton with Approximated Tensors (GNAT)
method of~\citet{Carlberg_JCP_2013}, have better stability properties, as they
formulate the ROM at the fully discrete level. The drawback of this approach is
that an additional layer of approximation -- usually called hyper-reduction ---
is required to gain computational speed-up. Moreover, the approach lacks
stability guarantees for low-dimensional expansions.

In this paper, a stabilization and enhancement approach to ROMs for the
compressible Navier-Stokes equations is developed.  The approach is an extension
of the methodology developed in~\cite{Balajewicz_ND_2012,Balajewicz_JFM_2013}
specifically for the incompressible Navier-Stokes equations. The specific volume
($\zeta$--) form of the compressible Navier-Stokes equations is utilized. Since
these equations have polynomial (quadratic) nonlinearities, the Galerkin
projection can be computed offline, once and for all; no hyper-reduction is
required~\cite{Iollo_TCFD_2000}. Unlike traditional eddy-viscosity-based
stabilization methods, the proposed approach requires no additional empirical
turbulence modeling terms -- truncated modes are accounted for \textit{a priori}
via a minimal rotation of projection subspace.  The method is also
non-intrusive, as it operates only on the matrices and tensors defining a ROM
ODE system, which are stabilized through the offline solution of a small
trace minimization problem on the Stiefel manifold. The proposed new approach can
be interpreted as a combination of several previously developed ideas.
Following~\citet{Iollo_TCFD_2000}, we propose to stabilize and enhance
projection-based ROMs by modifying the projection subspace in order to capture
more of the low-energy, but high dissipative scales of the flow solution.
Similarly to~\citet{Amsallem_IJNME_2012}, a rotation of the projection subspace is
used to achieve this goal.  Specifically, a larger set of basis is linearly
superimposed to provide a smaller set of basis that generate a stable and
accurate ROM. Finally, in the spirit of most previously proposed
eddy-viscosity-based turbulence models, the eigenvalues of the linear part of
the Galerkin system of ODEs are used as a proxy to guide the stabilization
algorithm.

The remainder of this paper is organized as follows. In \S~\ref{sect:meth}, the
standard projection-based model reduction approach is outlined in the context of
the specific volume form of the compressible Navier-Stokes equations.  Also
overviewed is the POD method for constructing an optimal reduced basis, and some
eddy-viscosity based closure models used for accounting for modal truncation.
The proposed methodology of ``rotating'' the projection subspace into a more
dissipative regime to better resolve the small, energy dissipative scales of the
flow is detailed in \S~\ref{sec:Proposed approach}. Here, the approach is
formulated mathematically as a constrained optimization problem on the Stiefel
manifold. In \S~\ref{sec:Applications}, the performance of the proposed method
is evaluated on several compressible flow problems, including a problem
involving a laminar flow over an airfoil with a high angle of attack, and a
channel-driven cavity flow problem.  Finally, conclusions are offered in
\S~\ref{sect:concl}.

\section{Projection-based model reduction for nonlinear compressible flow}
\label{sect:meth}

In this section the standard projection-based model reduction approach is laid
out. In \S~\ref{sec:Problem definition} the approach for a general nonlinear
system is presented while in \S~\ref{sec:ROMs_of_NS} the approach is applied to
the compressible Navier-Stokes equations.  The POD method for calculating a
reduced basis using a set of snapshots from a high-fidelity simulation is
outlined in \S~\ref{sect:POD}, followed by a brief overview of
eddy-viscosity-based closure models that account for modes truncated in the
application of the POD method (\S~ \ref{sect:closure}).

\subsection{Nonlinear projection-based model order reduction}
\label{sec:Problem definition}

Consider a dynamical system of the form:

\begin{equation}
\label{Eqn:EvolutionEquation}
    {\frac{d}{{dt}}\bm{w} } = \bm{F} \left( \bm{w} \right),
\end{equation}

\noindent where $\bm{F}$ is the propagator in $H$, a Hilbert
space. In fluid flows, the state variable
$\bm{w}=\bm{w}(\bm{x},t)\in H$ depends on space $\bm{x} \in \Omega$,
$\Omega$ being the flow domain, and time $t \in \left[ {0,T}
\right]$, $T$ representing the period of integration. Then, the
propagator $\bm{F}$ contains spatial derivatives. The associated
Hilbert space of square-integrable functions $L^2(\Omega)$ is
equipped with the standard inner product for its elements $ \bm{w}
\in L^2(\Omega)$, defined by:

\begin{equation}
    \left( {\bm{w},\bm{w}} \right)_\Omega  :=
    \int\limits_{\Omega} \bm{w} \cdot \bm{w} \> d\bm{x}.
\end{equation}

In the Galerkin ROM approach, the governing variable, $\bm{w}(\bm{x},t)$ is
discretized using basis functions (modes) $\left\{ \bm{w}_i (\bm{x})
\right\}_{i=1}^{n} \in H$ with corresponding mode coefficients $\left\{
a_i(t) \right\}_{i=1}^{n} $

\begin{equation}
\label{Eqn:GE}
    \bm{w}(\bm{x},t) \approx \bm{w}_0 (\bm{x}) + \bm{w}^{[1..n]} (\bm{w},t)
    := \bm{w}_0 (\bm{x}) + \sum\limits_{i=1}^n {a_i (t) \> \bm{w}_i
    (\bm{x})},
\end{equation}

\noindent where $\bm{w}_0(\bm{x})$ denotes the (steady) mean
flow.

In the method of lines, the modes $\bm{w}_i$ are known \textit{a
priori} and the goal is to find mode coefficients $a_i$ that satisfy
the differential equation \eqref{Eqn:EvolutionEquation}. In general,
the modes $\bm{w}_i$ can be chosen in a number of ways. In the
context of spectral methods in CFD for example, the basis vectors
are usually analytical functions, e.g.\ trigonometric functions or
Chebyshev polynomials. The advantage of these functions is that
their spatial derivatives have analytical representations and
numerically efficient algorithms such as the Fast Fourier Transform
(FFT) can be utilized. In the context of ROMs, the spatial basis
functions are usually derived \textit{a posteriori} from a snapshot
of a solution data set, like the Proper Orthogonal Decomposition
(POD)~\citep{Holmes_BOOK_2012} or Dynamic Mode Decomposition
(DMD)~\citep{Rowley_JFM_2009,Schmid_JFM_2010}. Attention is
restricted here to modes computed using the POD method (detailed in
$\S$~\ref{sect:POD}), but it is noted that the
methods proposed here hold for any choice of reduced basis. The
reason for the choice of the POD reduced basis is two-fold. First,
the POD is a widely used approach for computing efficient bases for
fluid dynamical systems. Moreover, ROMs constructed via the
POD/Galerkin method lack in general an \textit{a priori} stability
guarantee (meaning POD/Galerkin ROMs would benefit from ROM
stabilization approaches such as the one developed herein).

The mode coefficients in \eqref{Eqn:GE} $a_i$ are chosen
to minimize the residual of the Galerkin expansion

\begin{equation}
    \label{eqn:Galerkin_projection}
    \left( {\bm{w}_i ,{\frac{d}{{dt}}\bm{w}^{[1..n]} } } \right)_\Omega
    -\left( {\bm{w}_i , \bm{F} \left( \bm{w}_0 (\bm{x}) + {\bm{w}^{[1..n]} } \right)} \right)_\Omega   =
    0,
\end{equation}

\noindent for $i=1, ..., n$.  This projection yields a set
of evolution equations for the mode coefficients $a_i$

\begin{equation} \label{eq:evol}
    \frac{{d }}{{dt}}a_i  = f_i (\bm{a}),
\end{equation}

\noindent where $\bm{a} := (a_1,\ldots,a_n)^T$ represents the state
and $\bm{f} := (f_1,\ldots,f_n)^T$ its propagator. Given some
initial conditions, the evolution equation \eqref{eq:evol} can be
integrated using standard numerical integration techniques.
The ROM system \eqref{eq:evol}
is, by construction, small, and can be integrated forward in time in
real or near-real time unlike the full order dynamical system from
which it is derived.

\subsection{Nonlinear reduction of the compressible Navier-Stokes
equations} \label{sec:ROMs_of_NS}

Consider the 2D compressible Navier-Stokes equations in
primitive variables\footnote{Presented in two-dimensions for the
sake of brevity only. Extension to the three-dimensional equations
is straightforward.}:

\begin{subequations}
\label{eqn:NS_eq}
    \begin{align}
    \zeta_t + u \zeta_x + v \zeta_y -u_x \zeta - v_y \zeta &= 0, \\
    u_t + u u_x + v u_y + \zeta p_x &= \frac{1}{{\rm Re}} \zeta
        \left[
            \left(
                \frac{4}{3}u_x - \frac{2}{3}v_y
            \right)_x
            + (v_x + u_y)_y
        \right], \\
    v_t + u v_x + v v_y + \zeta p_y &= \frac{1}{{\rm Re}} \zeta
        \left[
            \left(
                \frac{4}{3}v_y - \frac{2}{3}u_x
            \right)_y
            + (v_x + u_y)_x
        \right], \\
    \begin{split}
    p_t + u p_x + v p_y + \gamma p (v_x + u_y) &=
        \frac{\gamma}{{\rm Re}\, {\rm Pr}}
        \left[
            (p \zeta)_{xx} + (p \zeta)_{yy}
        \right] \\
        &+ \frac{1-\gamma}{{\rm Re}}
        \left[
            u_x
            \left(
                \frac{4}{3} u_x - \frac{2}{3} v_y
            \right)
            + v_y
            \left(
                \frac{4}{3} v_y - \frac{2}{3} u_x
            \right)
            + (u_y +v_x)^2
        \right].
    \end{split}
    \end{align}
\end{subequations}
Here, $\zeta(\bm{x},t)=1/\rho(\bm{x},t)$ is the specific volume (the
inverse of the density, $\rho(\bm{x},t)$), $u(\bm{x},t)$
and $v(\bm{x},t)$ are the Cartesian components of
the flow velocity, $p(\bm{x},t)$ is the pressure, $\gamma$ is the
specific heat ratio, ${\rm Re}$ is the Reynolds number, ${\rm Pr}$
is the Prandtl number, and the subscripts denote
partial derivatives. A Galerkin projection yields a system of
coupled quadratic ODEs whose constant coefficients are calculated
off-line and once and for all (see~\ref{sec:Appendix_A}
and~\citet{Iollo_TCFD_2000} for details). This system has the form:

\begin{equation}
\label{eqn:Galerkin_system}
    \frac{d\bm{a}}{dt} = \bm{C} + \bm{L} \bm{a} +
        \left[ {\begin{array}{*{20}c} {\bm{a}^{\rm T} \bm{Q}^{(1)} \bm{a}}
        & {\bm{a}^{\rm T} \bm{Q}^{(2)} \bm{a}} &  \cdots
        & {\bm{a}^{\rm T} \bm{Q}^{(n)} \bm{a}}  \end{array} } \right]^{\rm
    T},
\end{equation}
where $\bm{C} \in \mathbb{R}^{n}$, $\bm{L} \in \mathbb{R}^{n \times
n}$ and $\bm{Q}^{(i)} \in \mathbb{R}^{n \times n}, \, \forall
i=1,\ldots,n$. \\

\noindent {\bf Remark 1:} The compressible Navier-Stokes equations
are typically expressed in conservative form. This form is
convenient for many applications including CFD.
The conservative form contains rational
functions of the unknowns and it is therefore not possible to
pre-compute ROMs using standard Galerkin projection; to attain any
computational speed-up a hyper-reduction step is necessary.
Hyper-reduction is not always desirable, as it can destroy energy
conservation properties and/or symplectic time-evolution maps
\cite{Carlberg_SIAM_2015, Kalashnikova_SAND_2014}.  On the other
hand, if the equations are expressed in primitive variables,
hyper-reduction can be avoided because all nonlinearities
that appear are polynomial.  It is
for this reason that, in our approach, we base the ROM on the
equations \eqref{eqn:NS_eq}. It is possible to extend our proposed
approach to problems where the full-order model (FOM) is based on
the compressible Navier-Stokes equations in conservative form; see
Remark 2.

\subsection{Construction of optimal reduced-order basis via the POD}
\label{sect:POD}

As discussed earlier, there exist a number of methods for
calculating a reduced basis $\left\{ \bm{w}_i (\bm{x})
\right\}_{i=1}^{n} \in H$, e.g., proper orthogonal decomposition
(POD)~\cite{Sirovich_QAP_1987, Holmes_BOOK_2012}, Dynamic Mode
Decomposition (DMD)~\citep{Rowley_JFM_2009,Schmid_JFM_2010} balanced
POD (BPOD)~\cite{Rowley_IJBC_2005, Willcox_AIAA_2002}, balanced
truncation~\cite{Gugercin_IJC_2004, Moore_IEEETAC_1981}, and the
reduced basis method \cite{Rozza_CCP_2011, Veroy_IJNMF_2005}. In
this paper, attention is restricted to reduced bases constructed
using the first of these approaches, namely the POD method.  This
method is reviewed succinctly below.

Discussed in detail in Lumley \cite{Lumley_Book_1971} and Holmes \textit{et al.}
\cite{Holmes_BOOK_2012}, POD is a mathematical procedure that, given an ensemble
of data and an inner product, constructs a basis for the ensemble.  The POD
basis is optimal in the sense that it describes more energy (on average) of the
ensemble in the chosen inner product than any other linear basis of the same
dimension $n$. Let $\bm{w}^n \in \mathbb{R}^{4N}$ denote a snapshot vector,
computed as the solution of the fully discretized version of
Eq.~\eqref{eqn:NS_eq}.  Here, $N$ is the number of grid points in the full order
model, for some instance of its parameters --- that is, for some specific time
$t$, some specific value of the set of flow parameters, or some boundary/initial
conditions underlying this governing equation. Suppose a total of $K \in
\mathbb{N}$ snapshots are collected from a high fidelity simulation. A snapshot
matrix is defined as a matrix $\bm{M} \in \mathbb{R}^{4N \times K}$ whose
columns are individual snapshots.  The main focus of this paper is on unsteady
flows and on snapshots associated with different time-instances. Hence, $M_{:,i}
:= \bm{w}^i$ for $i = 1,\ldots,K$. Mathematically, POD seeks an $n$-dimensional
($n \ll K$ and $n \ll 4N$) subspace spanned by the set $\left\{ \bm{w}_i (\bm{x})
\right\}_{i=1}^{n}$ such that the difference between the
ensemble $\{\bm{w}^i\}_{i=1}^K$ and its projection onto the reduced subspace is
minimized on average.  That is, a POD basis is obtained by solving the following
low-rank matrix approximation problem: \\

\noindent {\it For a given snapshot matrix}
$\bm{M}~\in~\mathbb{R}^{4N \times K}${\it , find a lower rank
matrix} $\widetilde{\bm{M}}~\in~\mathbb{R}^{4N \times K}$ {\it that
solves the minimization problem}

\begin{equation}
\label{eqn:LRA}
    \underset{{\rm rank}(\widetilde{\bm{M}})=n}{\text{min}}
    \displaystyle \| \bm{M} - \widetilde{\bm{M}} \|_F,
\end{equation}

\noindent \textit{where $n \ll 4N$.}\\

In this problem, the rank constraint can be taken care of by
representing the unknown matrix as $\widetilde{\bm{M}}=\widetilde{\bm{U}}
\widetilde{\bm{V}}$, where $\widetilde{\bm{U}}~\in~\mathbb{R}^{4N \times n}$ and
$\widetilde{\bm{V}}~\in~\mathbb{R}^{n \times K}$, so that
problem~(\ref{eqn:LRA}) becomes

\begin{equation}
\label{eqn:LRA2}
\underset{\widetilde{\bm{U}} \in \mathbb{R}^{4N \times n},\widetilde{\bm{V}} \in
\mathbb{R}^{n \times K}}{\text{min}}
\displaystyle \| \bm{M} - \widetilde{\bm{U}}\widetilde{\bm{V}} \|_F.
\end{equation}

\noindent It is well-known that the solution of the above low-rank
approximation problem is given by the
Eckart-Young-Mirsky~\cite{Eckart_P_1936,Mirsky_QJM_1960}
theorem via the
Singular Value Decomposition (SVD) of $\bm{M}$. Specifically,
$\widetilde{\bm{U}}=U_{:,1:n}$ and $\widetilde{\bm{V}}=(\bm{\Sigma}
\bm{V}^{{\rm T}})_{1:n,:}$ where $\bm{X}=\bm{U} \bm{\Sigma}
\bm{V}^{{\rm T}}$. This is the so-called ``method of snapshots" for
computing a POD basis~\cite{Sirovich_QAP_1987}.

\subsection{Accounting for modal truncation: eddy-viscosity based closure
models} \label{sect:closure}

In the Kolmogorov description of the turbulence cascade, the large,
energy--carrying flow scales transfer energy to successively smaller scales
where finally dissipative forces can dissipate their
energy~\cite{Tennekes_BOOK_1972,Pope_BOOK_2000}.  The large, energy-carrying
scales are associated with the large singular values of the snapshot matrix
$\bm{M}$, while the smaller, energy-dissipative scales of the flow are
associated with the smaller singular values.  Since low order POD-based ROMs
remove modes corresponding to small singular values, these ROMs are, by
construction, not endowed with the dissipative dynamics of the flow.

Many of the popular methods for accounting for truncated
modes fall in to the family of eddy-viscosity based closure models.
In this family of methods, dynamics of the truncated modes are
modeled by modifying the constant and linear terms of the Galerkin
model. In the most general case, the resulting modified Galerkin
system is of the following form:

\begin{equation}
\label{eqn:Galerkin_system_stab}
\frac{d\bm{a}}{dt} = \tilde{\bm{C}} + \tilde{\bm{L}}\bm{a}
    +
    \left[
        {\begin{array}{*{20}c} {\bm{a}^{\rm T} \bm{Q}^{(1)} \bm{a}}
                            & {\bm{a}^{\rm T} \bm{Q}^{(2)} \bm{a}}
                            &  \cdots
                            & {\bm{a}^{\rm T} \bm{Q}^{(n)} \bm{a}} \end{array}}
    \right]^{\rm T}.
\end{equation}

For a detailed review of the performance of the various methods based on this
approach the reader is referred to~\citet{Wang_CMAME_2012}.  Broadly speaking,
the modified constant and linear terms, $\tilde{\bm{C}}$, and $\tilde{\bm{L}}$,
are constructed in an effort to decrease energy production and increase energy
dissipation.  For example, consider the ``modal constant eddy-viscosity''
approach first proposed by~\citet{Rempfer_JFM_1994}. In this approach, the
linear term is modified using a constant matrix, $\tilde{\bm{L}} = \bm{L} +
\hat{\bm{L}}$ and the constant term is unmodified. This approach may be
interpreted as a modification of the eigenvalues of the linear operator. In
other words, the additional linear term is constructed to decrease the magnitude
of the real positive eigenvalues of $\bm{L}$ (i.e., decrease energy production)
and increase the magnitude of the negative eigenvalues of $\bm{L}$ (i.e.,
increase energy dissipation)\footnote{Other ROM stabilization approaches,
    developed independently from the ``modal constant eddy-viscosity'' approach
    and for a broader range of applications than fluid mechanics, e.g., the
    method of Kalashnikova et al.  \cite{Kalashnikova_CMAME_2014} for
stabilizing generic linear ROMs via optimization-based eigenvalue reassignment,
give rise to a similar modification to $\bm{L}$.}. Since the appropriate
eigenvalue distribution is not known {\it a priori}, the additional linear term
must be identified by solution matching techniques.

The linear eddy-viscosity approach has been successfully applied to a large
number of Galerkin models of complex, high-Reynolds number flows.  The method
has been particularly successful for ROMs of the incompressible Navier-Stokes
equations where energy conservation and symmetry arguments can be used to
further motivate the approach.

The main drawback of the eddy-viscosity based stabilization approach is the loss
of consistency between the Navier-Stokes equations and the Galerkin system. The
Galerkin system is modified empirically and, therefore, the quadratic system of
ODEs no longer corresponds to a Galerkin projection of the Navier-Stokes
equations. In the following section, a novel stabilization and enhancement
approach that retains consistency is introduced.

\section{Stabilization and enhancement of compressible flow ROMs via subspace
rotation}
\label{sec:Proposed approach}

In the new approach proposed herein, the truncated modes are modeled
{\it a priori} by ``rotating'' the projection subspace into
a more dissipative regime. Specifically, instead of
approximating the solution using only the first $n$ most energetic
POD modes, the solution is approximated using a linear-superposition
of $n+p$ (with $p>0$) most energetic POD modes. Mathematically this
can be expressed as:

\begin{equation}
    \label{eqn:superposition}
    \tilde{{\bm{w}}}_i  = \sum\limits_{j = 1}^{n+p} {X_{ji} } {{\bm{w}}_j}
    \hspace{5mm} i=1,\cdots,n,
\end{equation}

\noindent where $\mathbf{X} \in \mathbb{R}^{(n+p) \times n}$ is the
orthonormal ($\mathbf{X}^{T} \mathbf{X} = \mathbf{I}_{n \times n}$)
``rotation'' matrix. The Galerkin system tensors associated with
these new modes are expressed as a function of $\bm{X}$ as follows:

\begin{subequations}
    \label{eqn:new_system_from_X}
    \begin{align}
        \tilde{{Q}}_{jk}^{(i)} &=
        \sum\limits_{s,q,r = 1}^{n+p} {X_{si} Q_{qr}^{(s)} X_{qj}  X_{rk} } \hspace{5mm} i,j,k=1,\cdots,n,  \\
        \tilde{\bm{L}} &= \bm{X}^{\rm{T}} \bm{L} \bm{X}, \\
        \tilde{\bm{C}} &= \bm{X}^T \bm{C}^*,
    \end{align}
\end{subequations}

\noindent where ${\bm C} \in \mathbb{R}^{n+p}$, ${\bm L} \in
\mathbb{R}^{(n+p)\times (n+p)}$ and ${\bm Q}^{(i)} \in
\mathbb{R}^{(n+p)\times(n+p)}$,$\forall i=1,\cdots,(n+p)$ are the
Galerkin system coefficients corresponding to the first $n+p$ most
energetic POD modes.  The new Galerkin system is hence:

\begin{equation} \label{eqn:Galerkin_system_stab2}
\frac{d\bm{a}}{dt} = \tilde{\bm{C}} + \tilde{\bm{L}}\bm{a}
    +
    \left[
        {\begin{array}{*{20}c} {\bm{a}^{\rm T} \tilde{\bm{Q}}^{(1)} \bm{a}}
                            & {\bm{a}^{\rm T} \tilde{\bm{Q}}^{(2)} \bm{a}}
                            &  \cdots
                            & {\bm{a}^{\rm T} \tilde{\bm{Q}}^{(n)} \bm{a}} \end{array}}
    \right]^{\rm T},
\end{equation}

\noindent where the matrices $\tilde{\bm{Q}}^{(i)}$,
$\tilde{\bm{L}}$ and $\tilde{\bm{C}}$ are given by
\eqref{eqn:new_system_from_X}.

The goal of the proposed approach is to find $\bm{X}$ such that 1.) the new
modes $\tilde{{\bm{w}}}_i$ remain good approximations of the flow, and 2.) the
new Galerkin ROM is stable and accurate. To ensure that these properties are
satisfied, a constrained optimization problem is formulated for $\bm{X}$. To
guarantee that the new modes remain good approximations of the flow, the
distance $|| \bm{X} - \bm{I}_{n+p,n} ||_F$ is minimized, where $\bm{I}_{n+p,n}$
are the first $n$ columns of an ${n+p}$ identity matrix, and $||\cdot||_F$ is
the Frobenious norm To ensure that the ROM is stable and accurate, an
eddy-viscosity-based constraint equation is used as a proxy. In other words, the
eigenvalues of the linear operator are modified until a stable and accurate ROM
is generated.

Mathematically, the constrained optimization problem for
$\bm{X}$ outlined above reads as follows:

\begin{equation}
    \label{eqn:constrained low-rank problem}
    \begin{aligned}
    & \underset{\bm{X} \in \mathcal{V}_{(n+p),n}}{\text{minimize}}
        & & \displaystyle -{\rm{tr}}\left(\bm{X}^{\rm{T}}\bm{I}_{(n+p) \times n}\right)\\
        & \text{subject to}
            & & {\rm{tr}}( \bm{X}^{\rm{T}} \bm{L} \bm{X})=\eta
    \end{aligned}
\end{equation}
where $\eta \in \mathbb{R}$ and
\begin{equation} \label{eqn:Stiefel}
    \mathcal{V}_{(n+p),n} \in \{\bm{X}\in\mathbb{R}^{(n+p) \times n}~:~\bm{X}^{\rm T}
    \bm{X}=\bm{I}_n~,~p > 0\}.
\end{equation}
In \eqref{eqn:Stiefel}, $\mathcal{V}_{(n+p),n}$ is the Stiefel manifold,
defined as the set of $(n+p) \times n$ matrices satisfying the
orthonormality condition $\bm{X}^{\rm T}
\bm{X}=\bm{I}_n$~\cite{Rapcsak_EJOR_2002,Stiefel_CMH_1935}.  In
Equation~\eqref{eqn:constrained low-rank problem} the objective
function is simplified by utilizing the property that for a real
matrix $||\bm{A}||_F^2={\rm{tr}}(\bm{A}^{\rm T} \bm{A})$. Thus,
minimizing $|| \bm{X} - \bm{I}_{n+p,n} ||_F$ is equivalent to
minimizing $-{\rm{tr}}(\bm{X}^{\rm{T}}\bm{I}_{(n+p) \times n})$.

The appropriate eigenvalue distribution $\eta$, must be identified
using a solution matching procedure.  Discussion of an approach for
selecting $\eta$ is deferred until \S~\ref{sec:solution_matching}.\\

\noindent {\bf Remark 2:} In this paper, we assume that the ROM
advanced forward in time during the online time-integration step of
the model reduction is a system of the form
\eqref{eqn:Galerkin_system_stab}, which arises when projecting the
compressible Navier-Stokes equations in primitive specific volume
form \eqref{eqn:NS_eq} onto the reduced basis modes.
As suggested in Remark 1, the method described here can be applied
in the case the ROM is based on the compressible Navier-Stokes
equations in conservative form (with or without hyper-reduction). In
this case, the model reduction would proceed as follows:
\begin{itemize}
\item[\textit{Step 1:}] Run a high-fidelity code to generate
snapshots from which the POD basis will be constructed.
\item[\textit{Step 2:}] Construct from the snapshots collected in
Step 1 a POD basis $\left\{ \bm{w}_i (\bm{x})
\right\}_{i=1}^{n+p}$  for the primitive variables.
\item[\textit{Step 3:}] Project the compressible Navier-Stokes
equations in primitive specific-volume form \eqref{eqn:NS_eq} onto
the modes from Step 2 to obtain a system of the form
\eqref{eqn:Galerkin_system_stab}.
\item[\textit{Step 4:}] Use the matrices $\mathbf{L}$ and
$\mathbf{Q}^{(i)}$ to obtain from the original basis $\left\{ \bm{w}_i (\bm{x})
\right\}_{i=1}^{n+p}$ a
stabilized basis  $\tilde{{\bm{w}}}_i  = \sum\nolimits_{j = 1}^{n+p} {X_{ji} }
{{\bm{w}}_j},i=1,\cdots,n$, where ${{\bm X}}$ is the
solution to \eqref{eqn:constrained low-rank problem}.
\item[\textit{Step 5:}] Transform the stabilized basis $\left\{ \tilde{\bm{w}}_i (\bm{x})
\right\}_{i=1}^{n}$ into conservative variables, and use it in a ROM code that
projects the compressible Navier-Stokes equations in conservative
form (with or without hyper-reduction).
\end{itemize}
To apply this procedure, two ROM codes are required: a ROM code that
projects the compressible Navier-Stokes equations in primitive
specific-volume form, and a ROM code that projects the compressible
Navier-Stokes equations in conservative form.  The former code is
only needed to calculate the $\mathbf{L}$ and
$\mathbf{Q}^{(i)}$ matrices, which are used for the basis
stabilization.

\subsection{Solution of constrained optimization problem}

A common method for solving constrained optimization problems
of the form~\eqref{eqn:constrained low-rank problem} is the method
of Lagrange multipliers \cite{Nocedal_BOOK_1999}.  In this method,
the Lagrangian of the optimization problem is computed, and its
stationary points are sought, yielding necessary optimality
conditions for local maxima and minima.  The reader can verify that
the Lagrangian for Eq. \eqref{eqn:constrained low-rank problem} is
    \begin{equation}
        \mathcal{L}(\bm{X},\bm{\Lambda}_1,\bm{\Lambda}_2) :=
        -{\rm{tr}}(\bm{X}^{\rm{T}}\bm{I}_{(n+p) \times n})
        + {\rm{tr}}( \bm{\Lambda}_1(\bm{X}^{\rm T} \bm{L}^* \bm{X}-\tfrac{\eta}{n}\bm{I}_n))
        + {\rm tr} ( \bm{\Lambda}_2(\bm{X}^{\rm T} \bm{X} -
        \bm{I}_{n})),
    \end{equation}
where $\bm{\Lambda}_1$ and $\bm{\Lambda}_2$ are diagonal
matrices of Lagrange multipliers.

Suppose that $\bm{X}$ is a local maximizer of problem \eqref{eqn:constrained
low-rank problem}. Then $\bm{X}$ satisfies the first-order optimality condition
$\mathcal{L}_{\bm{X}}= - \bm{I}_{(n+p) \times n} + \bm{\Lambda}_1(\bm{L}^* +
{\bm{L}^*}^{\rm T}) \bm{X} +2\bm{\Lambda}_2 \bm{X} = 0$,
${\rm{tr}}(\bm{X}^{\rm{T}} \bm{L}^* \bm{X}-\tfrac{\eta}{n}\bm{I}_n) = \bm{0}$,
and $\bm{X}^{\rm T} \bm{X} - \bm{I}_{n} = \bm{0}$.  Solving
Eq.\eqref{eqn:constrained low-rank problem} using Lagrange multipliers is
possible; however, it is inefficient. Significant speed-ups are possible by
satisfying the orthonormality constraint directly via optimization on the
Stiefel matrix manifold.  In this method, with the help of the augmented
Lagrange method, the constrained optimization problem is reduced to an
unconstrained optimization problem on the Stiefel manifold as follows:

\begin{equation}
     \label{eqn:low-rank problem on Stiefel}
     \begin{aligned}
     & \underset{\bm{X} \in \mathcal{V}_{(n+p),n} }{\text{minimize}}
         & & \displaystyle -{\rm{tr}}( \bm{X}^{\rm{T}}\bm{I}_{(n+p) \times n})
         +\frac{\mu_k}{2}{\rm{tr}}(\bm{X}^{\rm T} \bm{L}^*
         \bm{X}-\tfrac{\eta}{n}\bm{I}_n)^2 - \lambda_{\mathcal{L}} {\rm{tr}}(\bm{X}^{\rm T} \bm{L}^*
         \bm{X}-\tfrac{\eta}{n}\bm{I}_n),
     \end{aligned}
 \end{equation}

\noindent where $\mu_k$ is increased until the constraint is
satisfied to some desired precision. The variable
$\lambda_{\mathcal{L}}$ is an estimate of the Lagrange multiplier
and is updated according to the rule

\begin{equation}
    {\lambda_{\mathcal{L}}}_{i+1} \leftarrow  {\lambda_{\mathcal{L}}}_i
    - \mu_k{\rm{tr}}{(\bm{X}^{(k)}}^{\rm T} \bm{L}^*
            \bm{X}^{(k)}-\tfrac{\eta}{n}\bm{I}_n),
\end{equation}

\noindent where $\bm{X}^{(k)}$ is the solution of the unconstrained problem at
the $k^{th}$ step. In this work, the {\tt Manopt} MATLAB
toolbox~\citep{Boumal_JMLR_2014} is used to solve~\eqref{eqn:low-rank problem on
Stiefel}. All derivatives in the optimization algorithm are calculated
analytically.

\subsection{Solution matching procedure for identification of stabilizing
eigenvalue spectrum $\eta$}
\label{sec:solution_matching}

In this work, the distribution of stabilizing eigenvalues
 $\eta$ in~\eqref{eqn:low-rank problem on
Stiefel} is identified using a brute force approach.  A brute force
approach is feasible in this case because the associated ROMs are
small and relatively inexpensive to integrate numerically.

Given the initial guess $\eta^{(0)}$, the optimization problem is
solved and the corresponding transformation matrix $\bm{X}^{(0)}$
identified. Using Eq.~\eqref{eqn:new_system_from_X} the new Galerkin
ROM is constructed and integrated numerically. A global ROM
``energy'' is defined as follows: $E(t)^{(k)}=\sum_i^n
(a(t)_i^{(k)})^2$. A linear fit of the temporal data is performed:
${E}(t)^{(k)} \approx c_1^{(k)} t + c_0^{(k)}$.  If the absolute
value of $c_1^{(k)}$ is below some predefined tolerance, the ROM is
deemed stable and the procedure halts. If the value of $c_1^{(k)}$
is positive (indicating increasing energy) then $\eta^{(k+1)} =
\eta^{(k)} + \epsilon$, where $\epsilon<0$.  If the value of
$c_1^{(k)}$ is negative (indicating decreasing energy) then
$\eta^{(k+1)} = \eta^{(k)} + \epsilon$ where $\epsilon>0$. This
search is automated using MATLAB's \verb=fzero= root finding
algorithm. The overall stabilization procedure is summarized in
Algorithm~\ref{alg:stabilization}.  In Remark 3 below, we
give some guidelines for selecting the parameters in
Algorithm~\ref{alg:stabilization}, and some general remarks about
the numerical solution of the
optimization problem~\eqref{eqn:constrained low-rank problem}.\\

\noindent {\bf Remark 3.} Here, we provide some general
guidelines and remarks pertaining to Algorithm 1.
\begin{itemize}
\item We recommend initializing the algorithm with $\bm{X}^{(0)}=\bm{I}_n$
    and $\eta^{(0)} = (\bm{X}^{(0)}\bm{L}\bm{X}^{(0)}) = tr(\bm{L}_{1:n,1:n})$,
    which corresponds to the trivial solution.

\item Our numerical experiments suggests $n=p$ provides best performance; however
     the optimal choice of $p$ remains an open question.

\item The proposed algorithm requires that $p\geq1$.
    For $p=0$, the rotation matrix $\bm{X}$ is square and therefore
    corresponds to a coordinate transformation that, by definition,
    can have no effect on the dynamics of the system.

\item Existence and uniqueness results for the solution to
    \eqref{eqn:constrained low-rank problem} are not provided in this work.  We
    have found the results to be insensitive to the initial conditions, and the
    solutions to \eqref{eqn:constrained low-rank problem} to be unique for a
    large number of random initial conditions.

\item For some choices of $p$ and $\eta$ the optimization problem~\eqref{eqn:low-rank problem on
    Stiefel} has no solutions.
    Let $\lambda_1>\lambda_2>\cdots>\lambda_{n+p}$ be the
    eigenvalues of the symmetric part of $\bm{L}$ and let $\bm{V}$ be a matrix of
    the associated eigenvectors\footnote{We can look at the symmetric part
        with out loss of generality since
    ${\rm{tr}}(\bm{L}) = {\rm{tr}}((\bm{L}+\bm{L}^T)/2)$}.  By definition
    ${\rm{tr}}(\bm{L}) = \sum_{i=1}^{n+p} \lambda_i$.
    Therefore, $\sum_{i=1}^n(\lambda_i) \leq \eta \leq
    \sum_{i=n+1}^{n+p}(\lambda_i)$ is a necessary condition for the existence of a
    solution to~\eqref{eqn:low-rank problem on
    Stiefel}. The lower and upper bounds for this $\eta$ are
    provided by the transformations $\bm{X}=\bm{V}_{:,1:n}$ and
    $\bm{X}=\bm{V}_{:,n+1:n+p}$, respectively.
\end{itemize}

\begin{algorithm}
\caption{Stabilization algorithm for compressible Navier-Stokes equations
\label{alg:stabilization}}
\DontPrintSemicolon
\SetKwInOut{Input}{input}
\SetKwInOut{Output}{output}
\Input
    {Initial guess for stabilizing eigenvalue distribution $\eta^{(0)}$, ROM
    size $n$ and $p\geq1$,
    Galerkin system matrices associated with the first $n+p$ most
    energetic POD modes.
    ${\bf C} \in \mathbb{R}^{n+p}$,
    ${\bf L} \in \mathbb{R}^{(n+p) \times (n+p)}$ and
    ${\bf Q}^{(i)} \in \mathbb{R}^{(n+p) \times (n+p)}, i=1,\cdots,n+p$,
    convergence tolerance $TOL$, and maximum number of
    iterations.
    $k_{max}$
    }
\Output{Stabilizing rotation matrix $\bm{X}$}
\BlankLine
\For{$k=0,\cdots,k_{max}$}
{
    Solve constrained optimization problem on Stiefel
            manifold:
            \begin{equation*}
                \begin{aligned}
                    & \underset{\bm{X}^{(k)} \in \mathcal{V}_{(n+p),n}}{\text{minimize}}
                    & & \displaystyle -{\rm{tr}}\left(\bm{X}^{(k)\rm{T}} \bm{I}_{(n+p) \times n}\right)\\
                    & \text{subject to}
                    & & {\rm{tr}}( \bm{X}^{(k)\rm{T}} \bm{L}
                    \bm{X}^{(k)})=\eta^{(k)}.
                \end{aligned}
            \end{equation*}\;

    Construct new Galerkin matrices using~\eqref{eqn:new_system_from_X}\;

    Integrate numerically new Galerkin system\;
    Calculate modal energy $E(t)^{(k)}=\sum_i^n (a(t)_i^{(k)})^2$\;
    Perform linear fit of temporal data ${E}(t)^{(k)} \approx c_1^{(k)}t+c_0^{(k)}$\;
    Based on energy growth $c_1^{(k)}$, calculate $\epsilon$ using root-finding algorithm and
    perform update $\eta^{(k+1)} = \eta^{(k)} + \epsilon$\;
    \If{$||c_1^{(k)}||<TOL$}
    {
        $\bm{X}:=\bm{X}^{(k)}$\;
        \textbf{terminate the algorithm}\;
    }
}
\end{algorithm}

\section{Numerical experiments}
\label{sec:Applications} In this section, we evaluate the
performance of Algorithm \ref{alg:stabilization} for stabilizing
compressible flow ROMs on several 2D problems:
a problem involving a laminar flow around an inclined
airfoil, and a channel driven cavity problem at two Reynolds
numbers. In all cases, the flow is governed by the
full compressible Navier-Stokes equations with
constant viscosity. Direct Numerical Simulations (DNS) are performed
and POD basis functions are derived from snapshots collected during
these simulations.  ROMs are derived by
projecting the
fully compressible Navier-Stokes equations in specific volume
($\zeta$--) form onto the first $n$ most energetic basis
modes. The projection is performed off-line, and
once and for all, resulting in a system of $n$ coupled quadratic
ODEs in the form of~\eqref{eqn:Galerkin_system}.
 From this point forward, such ROMs are
referred to as ``standard POD-Galerkin ROMs", where ``standard"
refers to the fact that no model is used to account for the dynamics
of the truncated modes, $n+1,n+2,\cdots,\infty$. ROMs derived using
the stabilization approach proposed in this paper are referred to as
``stabilized POD-Galerkin ROMs".

\subsection{High angle of attack laminar airfoil}
The first test case involves the 2D flow around an inclined NACA0012
airfoil at Mach 0.7, and ${\rm Re}=500$ at $25$
degrees angle of attack. The Reynolds number is based on the chord
of the airfoil, $c$.  At this Reynolds number and angle of attack,
the flow is separated and the solution corresponds to a stable limit
cycle. High-fidelity simulation snapshots are generated using a
second-order, finite-difference, embedded boundary (EB) solver. The
no-slip adiabatic boundary conditions are satisfied using a
first-order ghost fluid method. For a detailed description of the
scheme the reader is referred to~\citet{Balajewicz_JCP_2014}. The
snapshots correspond to a DNS of the compressible
Navier-Stokes equations. The viscosity of the fluid inside the
domain is assumed constant. The computational domain extends $20c$
in all directions and a sponge zone of thickness $2c$ is used to
absorb all waves entering and leaving the domain.  The domain is
discretized using a non-uniform $300\times300$ tensor grid.  The
flow is initialized by setting the solution at all grid points to the
free stream values. Time integration is performed using an implicit
second order BDF scheme and a constant time step
corresponding to a CFL=$1$ is used. Snapshot collection begins after
$5000$ time steps to ensure the solution has reached the limit
cycle. A total of $K=500$ snapshots are collected every 5 simulation
time steps. The first four basis functions capture approximately 86\% of the snapshot
energy.

\begin{figure}
\centering
\subfigure[\label{fig:airfoil_E}]
{
    \includegraphics{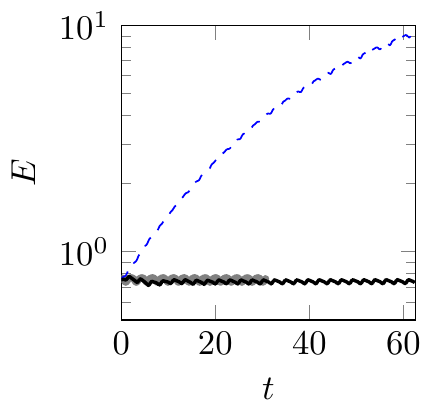}
}
\hspace{-0.2cm}
\subfigure[\label{fig:airfoil_a1a2} ]
{
    \includegraphics{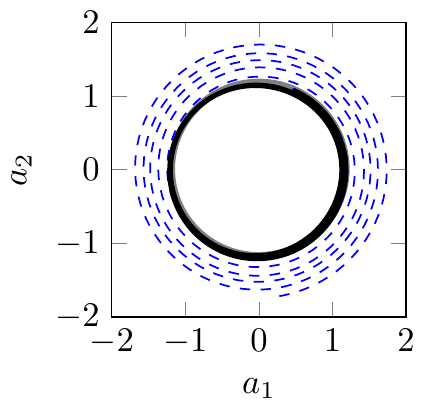}
}
\hspace{-0.2cm}
\subfigure[\label{fig:airfoil_X} ]
{
    \includegraphics{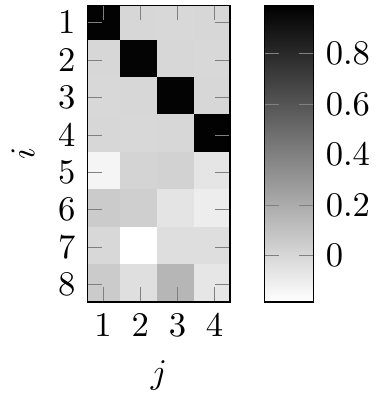}
}
\caption
    {
    Nonlinear model reduction of the laminar airfoil. Evolution of modal energy
    (a), and phase plot of the first and second temporal basis, $a_1(t)$ and
    $a_2(t)$ (b); DNS (thick gray line),
    standard $n=4$ ROM (dashed blue line),
    stabilized $n,p=4$ ROM (solid black line).
    Stabilizing rotation matrix, $\bm{X}$ (c)
    }\label{fig:airfoil}
\end{figure}

Figure~\ref{fig:airfoil} illustrates the performance of a stabilized
$n,p=4$ ROM of the laminar airfoil. In Figure~\ref{fig:airfoil_E},
the global energy of the DNS, standard (i.e., unstabilized) and
stabilized ROMs are illustrated. The stabilized ROM is shown to
track very accurately the mean of the fluctuating energy of the
original DNS snapshots while the standard ROM over predicts
the mean by an order of magnitude. To investigate the long term
stability of the stabilized ROM, the system was numerically integrated
$100\times$ the duration of the original snapshots. No change or drift in trajectory
was observed during this long integration period. In
Figure~\ref{fig:airfoil_a1a2} the trajectories of the first and
second temporal coefficient, $a_1(t),$ and $a_2(t)$ respectively,
are illustrated. The stabilized ROM accurately reproduces
the closed orbit of the stable limit cycle while the standard ROM
predicts an unstable spiral.
Figures~\ref{fig:airfoil_E} and \ref{fig:airfoil_a1a2} demonstrate
how the stabilized ROM reproduces reliably both the global mean and
fluctuating components of the DNS snapshots.
The stabilizing transformation matrix
$\bm{X}$ for this problem is illustrated Figure~\ref{fig:airfoil_X}.
As expected the rotation of the projection subspace is small as
demonstrated by the fact that $\bm{X} \approx \bm{I}_{(n+p) \times
n}$. For this configuration, the normalized error defined as
$||\bm{X}-\bm{I}_{(n+p) \times n}||_F/n$ is $0.083$.

\begin{figure}
\centering

    \includegraphics{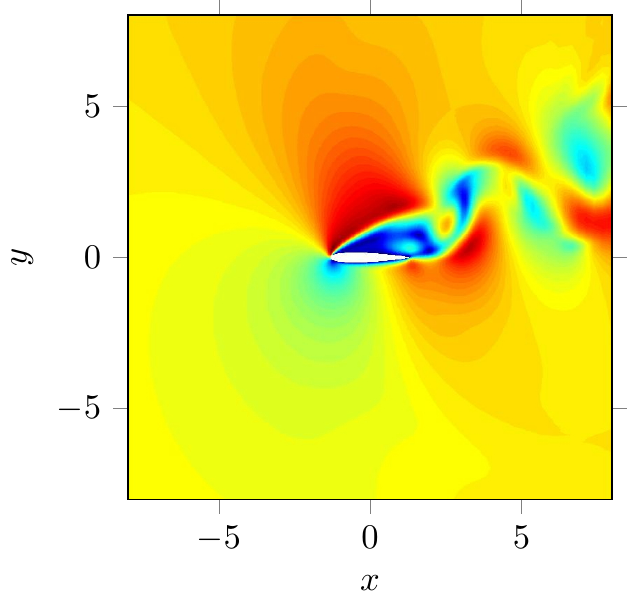}

    \includegraphics{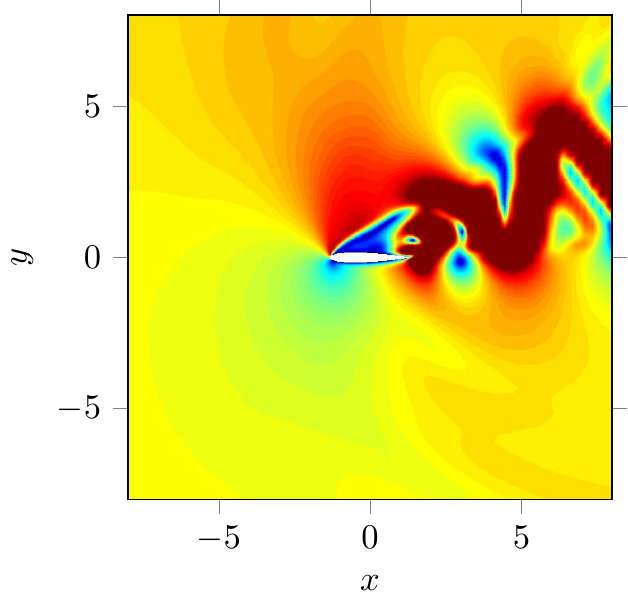}

    \includegraphics{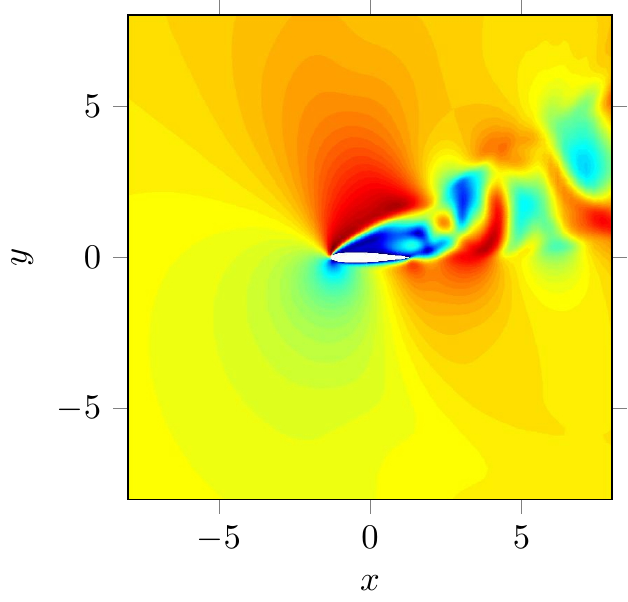}

\caption
    {
    Snapshot of high angle of attack airfoil at final snapshot;
    contours of velocity magnitude.
    DNS (top),
    standard $n=4$ ROM (middle),
    and stabilized $n,p=4$ ROM (bottom)
    }\label{fig:airfoil_snapshot}
\end{figure}

Finally, the predicted velocity magnitude at the final snapshot is
illustrated in Figure~\ref{fig:airfoil_snapshot}. The stabilized ROM
(Figure~\ref{fig:airfoil_snapshot}(a)) reproduces the
velocity contours of the original high-fidelity DNS simulation
(Figure~\ref{fig:airfoil_snapshot}(c)) remarkably
well, in contrast to the standard ROM
(Figure~\ref{fig:airfoil_snapshot}(b)). This demonstrates the
effectiveness of the proposed model reduction approach.


\subsection{Channel driven laminar cavity}
\label{sec:Channel driven laminar cavity} For the results presented
in this section, the high-fidelity fluid simulation data are
generated using a Sandia National Laboratories' in-house finite
volume flow solver known as SIGMA CFD. This code is derived from
LESLIE3D \cite{Sankaran_PCI_2005}, a Large Eddy Simulations (LES)
flow solver originally developed in the Computational Combustion
Laboratory at the Georgia Institute of Technology. The code has LES
as well as DNS capabilities.  For the channel driven laminar cavity
problem considered here, the code was run in DNS mode. For
a detailed description of the schemes and models implemented within
LESLIE3D, the reader is referred to \cite{Genin_JofT_2010,
Genin_CF_2010}.

ROMs for the channel driven laminar cavity problem are constructed
using a Sandia in-house parallel C++ model reduction code known as
{\tt Spirit}, which constructs ROMs for compressible flow problems
using the POD and continuous
projection method. This code, detailed in
\cite{Kalashnikova_SAND_2014}, reads in the snapshot and mesh data
written by a high-fidelity flow solver, creates a finite element
representation of the snapshots and computes the numerical
quadrature necessary for evaluation of the inner products arising in
the Galerkin projection step of the model reduction. All
calculations are performed in parallel using distributed matrix and
vector data structures and parallel eigensolvers from the Trilinos
project \cite{Heroux_ACMTMS_2005}, which allows for large data sets
and a relatively large number of POD modes. The {\tt libmesh} finite
element library \cite{Kirk_EC_2006} is used to compute the element
quadratures.

In the discussion that follows, two variants of the 2D channel
driven laminar cavity problem are considered: a low Reynolds number
variant (${\rm Re} \approx 1500$) and a moderate Reynolds number
variant (${\rm Re} \approx 5500$).  Both tests cases involve a Mach
0.6 viscous laminar flow over a cavity in a $T$--shaped domain
(Figure \ref{fig:cavity_mesh2}). The flow conditions for both tests
are similar to case L2 in \cite{Rowley_JFM_2002}.  The free stream
pressure is $25$ Pa, the free stream temperature is $300$ K, and the
free stream velocity is $208.8$ m/s.  The viscosity $\mu$ is
spatially constant and calculated such that the above Reynolds
numbers are achieved. The thermal conductivity $\kappa$ is also
constant, calculated such that the Prandtl number is Pr$ = 0.72$. At
the inflow boundary, a value of the velocity and temperature that is
above the free stream values is specified.  The flow at the cavity
walls is assumed to be adiabatic and to satisfy a no-slip condition.
The remaining outflow boundaries are open, and a far-field boundary
condition that suppresses the reflection of waves into the
computational domain is implemented here. The high-fidelity
simulation is initialized by setting the flow in the cavity to have
a zero velocity, free stream pressure, and temperature. The region
above the cavity is initialized to free stream conditions and the
flow is allowed to evolve.  As both SIGMA CFD and {\tt Spirit} are
3D codes, a 2D mesh of the domain $\Omega$ is converted to a 3D mesh
by extruding the 2D mesh in the $z$-direction by one element.
Finally, it is noted that SIGMA CFD and {\tt Spirit} work
with different meshes.  The former code requires a structured
hexahedral discretization, whereas the latter assumes a tetrahedral
discretization. To overcome this difference, the hexahedral
high-fidelity meshes associated with the snapshots are converted to
tetrahedral meshes prior to constructing the ROMs. This is
accomplished by breaking up each hexahedral element into six
tetrahedral elements.

\subsubsection{Low Reynolds number $({\rm Re} \approx 1500)$}
\label{sec:Re1500}

\begin{figure}
\label{fig:cavity_mesh2} \centering
    \includegraphics{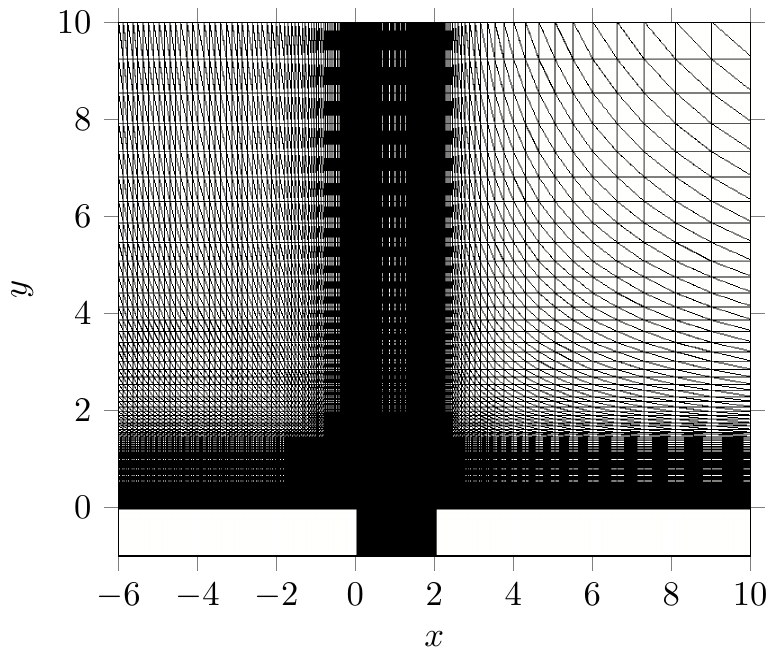}
\caption{Domain and mesh for viscous channel driven cavity problem}
\end{figure}

For the first, low Reynolds number variant of the channel driven
laminar cavity problem, the free-stream viscosity is set to
$\mu_{\infty} = 3.17\times 10^{-6}$ kg/(m$\cdot$s), so that ${\rm
Re} = 1453.9$. The discretized domain, illustrated in Figure
\ref{fig:cavity_mesh2}, consists of 98,408 nodes, cast as 292,500
tetrahedral finite elements within the ROM code, {\tt Spirit}.  The
2D extent of the domain is: $[(-6.42, 10) \times (-1,10)] \backslash
[(-6.42, 10)\times(-1,0) \cup (2, 10) \times (-1, 0)]$ m.   The
reader can observe that the mesh is structured but non-uniform.

The high-fidelity solver, SIGMA CFD, is initiated with the
conditions described above and allowed to run until a statistically
stationary flow regime is reached.  At this point, a total of
$K_{max} = 500$ snapshots are collected from SIGMA CFD, taken every
$\Delta t_{snap} = 1 \times 10^{-4}$ seconds. The snapshots are used
to construct a POD basis of size 4 modes in the $L^2$ inner product.
This basis captured about 91\% of the snapshot energy.
For more details on this test case, the reader is referred to
\cite{Kalashnikova_SAND_2014}.

\begin{figure}
\centering
\subfigure[\label{fig:low_Re_cavity_E}]
{
    \includegraphics{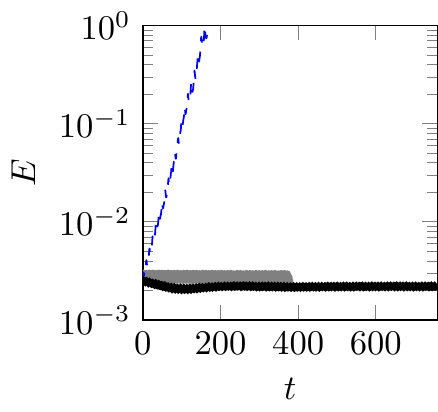}
}
\hspace{-0.2cm}
\subfigure[\label{fig:low_Re_cavity_a1a2} ]
{
    \includegraphics{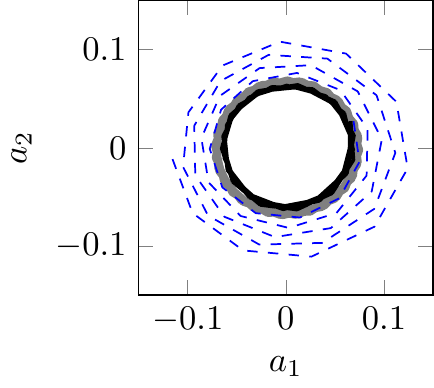}
}
\hspace{-0.2cm}
\subfigure[\label{fig:low_Re_cavity_X} ]
{
    \includegraphics{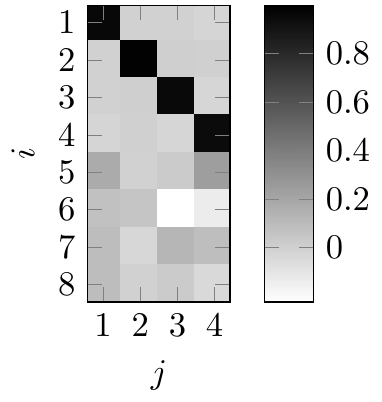}
}
\caption
    {
    Nonlinear model reduction of channel drive cavity at ${\rm Re} \approx 1500$.
    Evolution of modal energy (a) and phase plot of the first and second
    temporal basis, $a_1(t)$ and  $a_2(t)$ (b); DNS (thick gray line),
    standard $n=4$ ROM (dashed blue line), stabilized $n,p=4$ ROM (solid
    black line).  Stabilizing rotation matrix, $\bm{X}$ (c)
    }\label{fig:low_Re_cavity}
\end{figure}

Figure~\ref{fig:low_Re_cavity} illustrates the performance of a stabilized
$n,p=4$ ROM of the low Reynolds number cavity.  In
Figure~\ref{fig:low_Re_cavity_a1a2}, the modal energy of the DNS, standard, and
stabilized ROMs are illustrated. The stabilized ROM is shown to track very
accurately the energy of the original DNS snapshots while the standard ROM is
unable to reproduce this trajectory. The long term stability of the stabilized
ROM was validated by numerically integrating the system $100\times$ the duration
of the original snapshots. No change or drift in trajectory was observed during
this long integration period. In Figure~\ref{fig:low_Re_cavity_a1a2} the
trajectories of the first and second temporal coefficient, $a_1(t),$ and
$a_2(t)$ respectively, are illustrated. The stabilized ROM predicts correctly
the closed orbit of the stable limit cycle while the standard ROM predicts an
unstable spiral. The stabilizing transformation matrix $\bm{X}$ for this problem
is illustrated Figure~\ref{fig:low_Re_cavity_X}. As before, the rotation of the
projection subspace is small as demonstrated by the fact that $\bm{X} \approx
\bm{I}_{(n+p) \times n}$. For this configuration, the normalized error defined
as $||\bm{X}-\bm{I}_{(n+p) \times n}||_F/n$ is $0.1182$.

\begin{figure}
\centering

    \includegraphics{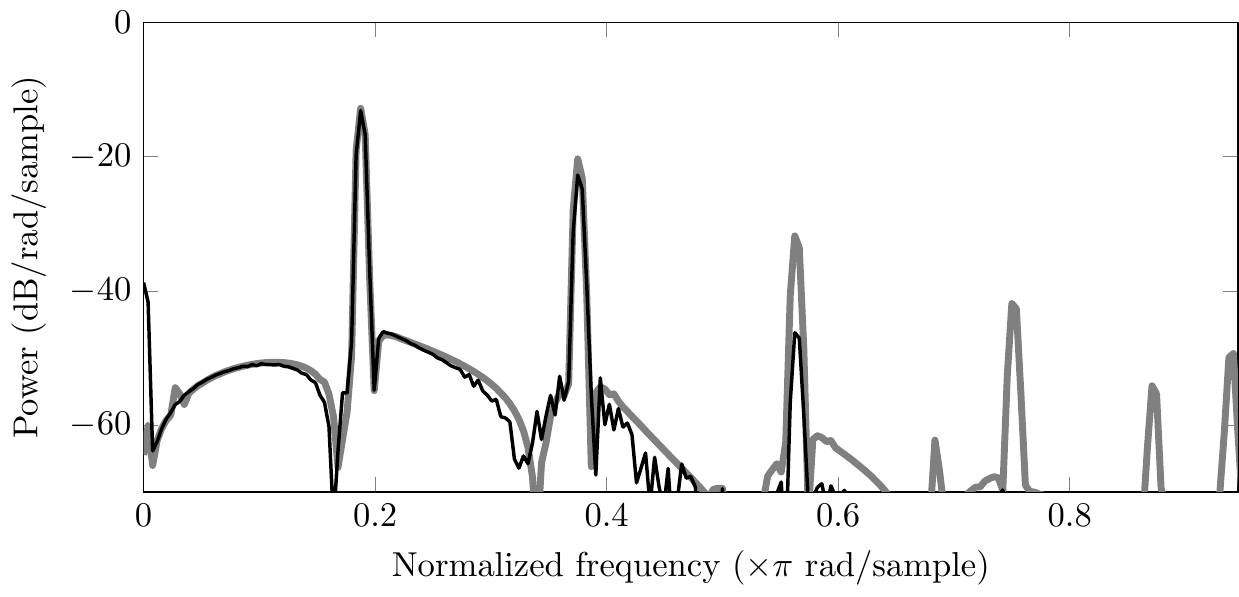}

\caption{
        PSD of $p(\bm{x},t)$ where $x=(2,-1)$ of
        channel drive cavity ${\rm Re} \approx 1500$.
        DNS (thick gray line),
        stabilized $n,p=4$ ROM (black line)
        }\label{fig:low_Re_cavity_PSD}
\end{figure}

Figure~\ref{fig:low_Re_cavity_PSD} shows the Power Spectral Density (PSD) of the
predicted pressure fluctuations at the bottom right corner of the cavity,
$\bm{x}=(2,-1)$, Both the fundamental and first harmonic of the response is
accurately predicted by the stabilized $n,p=4$ ROM. The PSD of the CFD signal
was computed using all available snapshots from $t=0$ to $t=380$ where $t$ is
non-dimensional. On the other hand, the PSD of the stabilized ROM was computed
from the signal $100\times$ past the duration of the original snapshots; i.e.
$t=(38000-380)$ to $t=38000$.

\begin{figure}
\centering

    \includegraphics{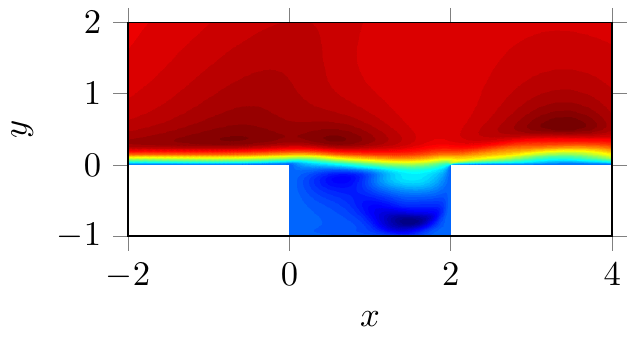}

    \includegraphics{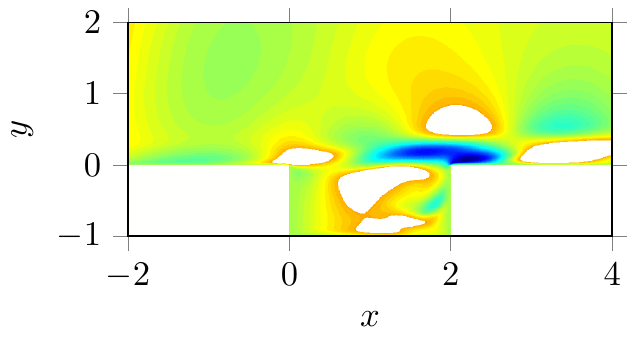}

    \includegraphics{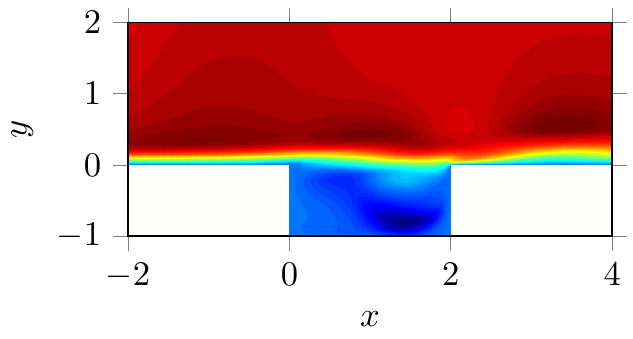}

\caption{
        Snapshot of channel drive cavity ${\rm Re} \approx 1500$;
        contours of $u$-velocity magnitude at the final snapshot.
        DNS (top),
        standard $n=4$ ROM (middle)
        and stabilized $n,p=4$ ROM (bottom)
        }\label{fig:low_Re_cavity_snapshot}
\end{figure}

Finally, a snapshot of the predicted
velocity magnitude at the final snapshot is illustrated in
Figure~\ref{fig:low_Re_cavity_snapshot}. The stabilized ROM
(Figure~\ref{fig:low_Re_cavity_snapshot}(a)) reproduces
the velocity contours of the original high-fidelity DNS simulation
(Figure~\ref{fig:low_Re_cavity_snapshot}(c)) remarkably
well.  In contrast, the standard ROM
(Figure~\ref{fig:low_Re_cavity_snapshot}(c)) is unstable and
inaccurate.

\subsubsection{Moderate Reynolds number $({\rm Re} \approx 5500)$}
\label{sec:Re5500}

The next test case considered is also a channel drive laminar cavity
problem, but at a higher Reynolds number.  The only parameter that
is different is the free-stream viscosity, now set to $\mu_{\infty}
= 8.46\times 10^{-7}$ kg/(m$\cdot$s), so that  ${\rm Re} = 5452.1$.  Also
changed is the size of the geometry extent, which has a larger
sponge region near the outflow regions.   This is needed to suppress
adequately the reflection of waves into the computational domain for
this problem.  Toward this effect, the 2D extent of the domain is:
$[(-6.42, 30) \times (-1,30)] \backslash [(-6.42, 30)\times(-1,0)
\cup (2, 30) \times (-1, 0)]$ m.  The geometry is discretized by
117,328 nodes, cast as 345,900 tetrahedral elements in {\tt Spirit}.
As before, the mesh is structured but non-uniform. The flow is
significantly more chaotic than the ${\rm Re} \approx 1500$ case
considered in Section \ref{sec:Re1500}.

A total of $K = 500$ snapshots are collected from SIGMA CFD at
increments $\Delta t_{snap} = 1 \times 10^{-5}$ seconds.  As before,
snapshots collection does not begin until a statistically stationary
flow regime has been reached.  From these snapshots, a POD basis of
size 20 modes is constructed in the $L^2$ inner product. This basis
captures about 72\% of the snapshot energy.  Typically, $n$
would be selected such that the POD basis captures a greater
percentage of the snapshot ensemble energy (e.g., $\approx 90\%$ or
more).  We choose a basis that captures less energy of the snapshot
set to highlight the effectiveness of our approach for
low-dimensional POD expansions.

\begin{figure}
\centering
\subfigure[\label{fig:high_Re_cavity_E}]
{
    \includegraphics{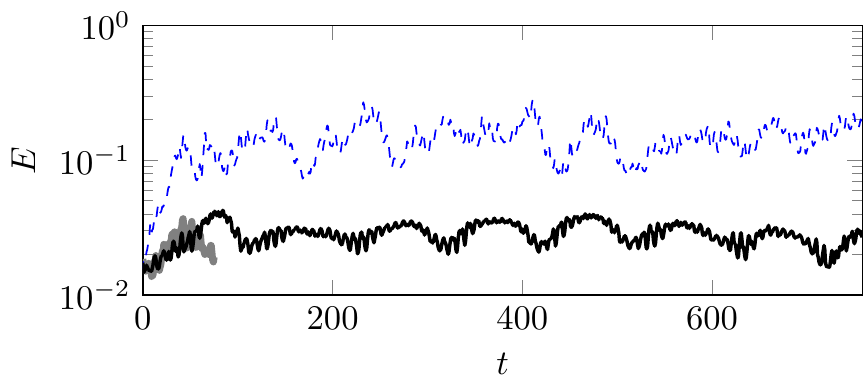}
}
\hspace{-0.2cm}
\subfigure[\label{fig:high_Re_cavity_X} ]
{
    \includegraphics{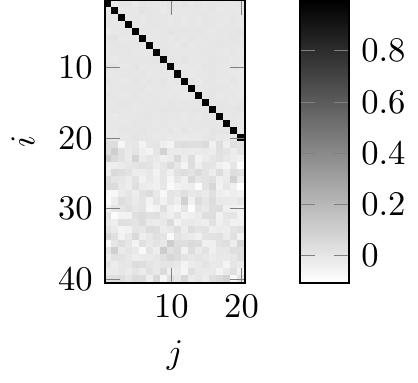}
}
\caption
    {
    Nonlinear model reduction of channel drive cavity at ${\rm Re} \approx 5500$.
    Evolution of modal energy (a); DNS (thick gray line),
    standard $n=20$ ROM (dashed blue line), stabilized $n,p=20$ ROM (solid
    black line). Stabilizing rotation matrix, $\bm{X}$
    (b)
    }\label{fig:high_Re_cavity}
\end{figure}

Figure~\ref{fig:high_Re_cavity} illustrates the performance of a
stabilized $n,p=20$ ROM of the higher Reynolds number cavity
problem. In Figure~\ref{fig:high_Re_cavity_E}, the modal energy of
the DNS, standard, and stabilized ROMs are illustrated. The standard
ROM is shown to over predict the energy of the original DNS
snapshots by an order of magnitude. The predictive power of the
stabilized ROM is demonstrated by numerically integrating the ROM
$10\times$ the duration of the original snapshots. The stabilizing
transformation matrix $\bm{X}$ for this problem is illustrated
Figure~\ref{fig:high_Re_cavity_X}. As before, the rotation of the
projection subspace is small as demonstrated by the fact that
$\bm{X} \approx \bm{I}_{(n+p) \times n}$. For this configuration,
the normalized error defined as $||\bm{X}-\bm{I}_{(n+p) \times
n}||_F/n$ is $0.0384$.

\begin{figure}
\centering
    \includegraphics{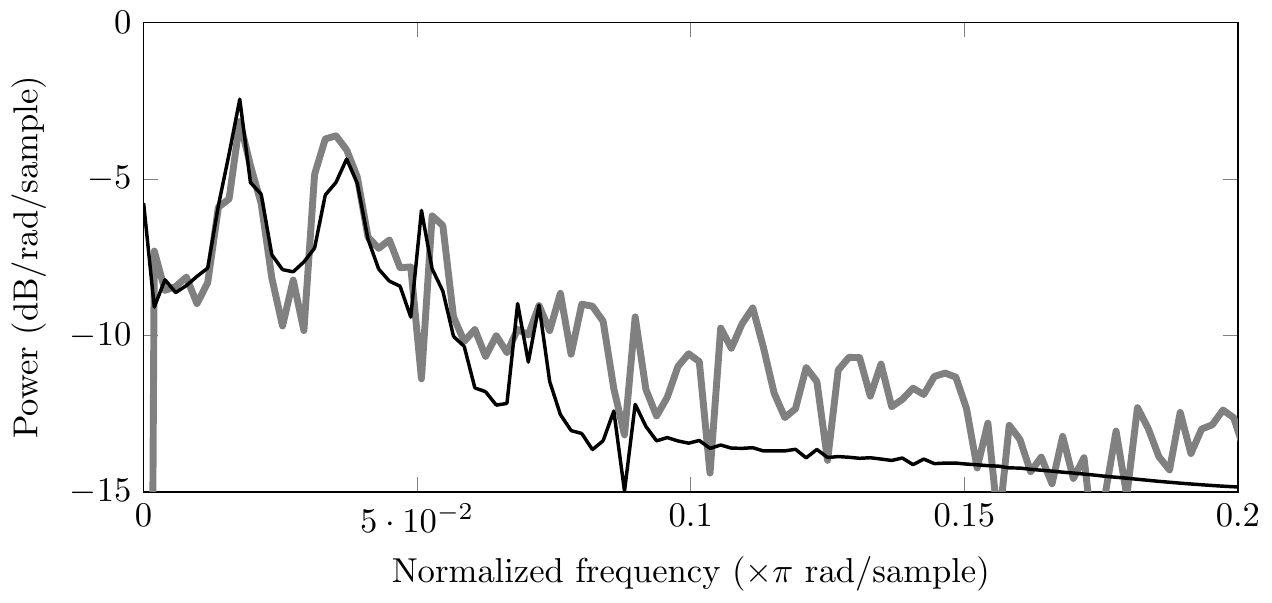}
\caption{
        PSD of $p(\bm{x}_1,t)$ where $\bm{x}_1 = (2,-0.5)$
        of channel driven cavity at ${\rm Re} \approx 5500$.
        DNS (thick gray line),
        stabilized $n,p=20$ ROM (black line)
        }\label{fig:high_Re_cavity_PSD1}
\end{figure}

\begin{figure}
\centering
    \includegraphics{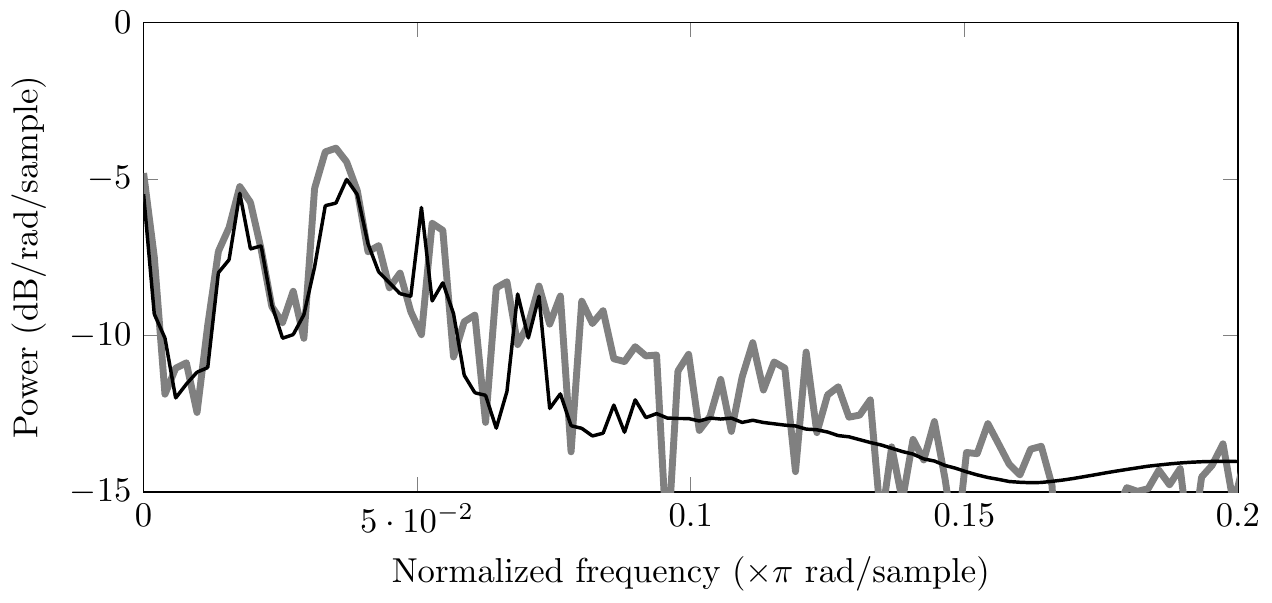}
\caption{
        PSD of $p(\bm{x}_2,t)$ where $\bm{x}_2 = (0,-0.5)$
        of channel driven cavity at ${\rm Re} \approx 5500$.
        DNS (thick gray line),
        stabilized $n,p=20$ ROM (black line)
        }\label{fig:high_Re_cavity_PSD2}
\end{figure}

\begin{figure}
\centering
    \includegraphics{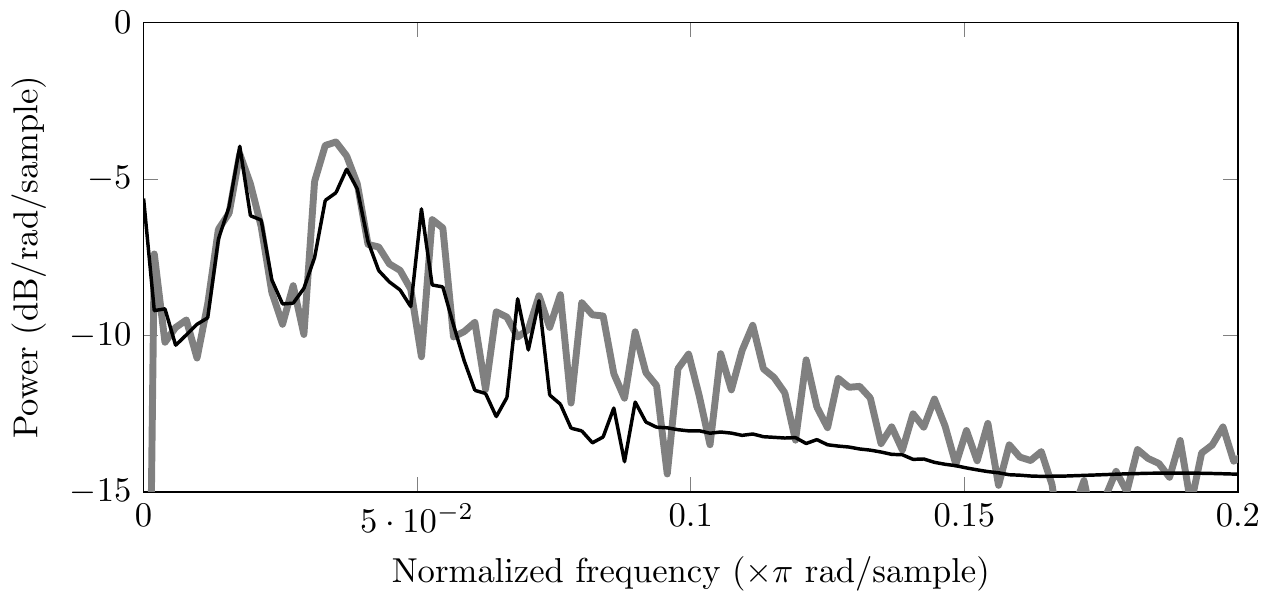}
    \includegraphics{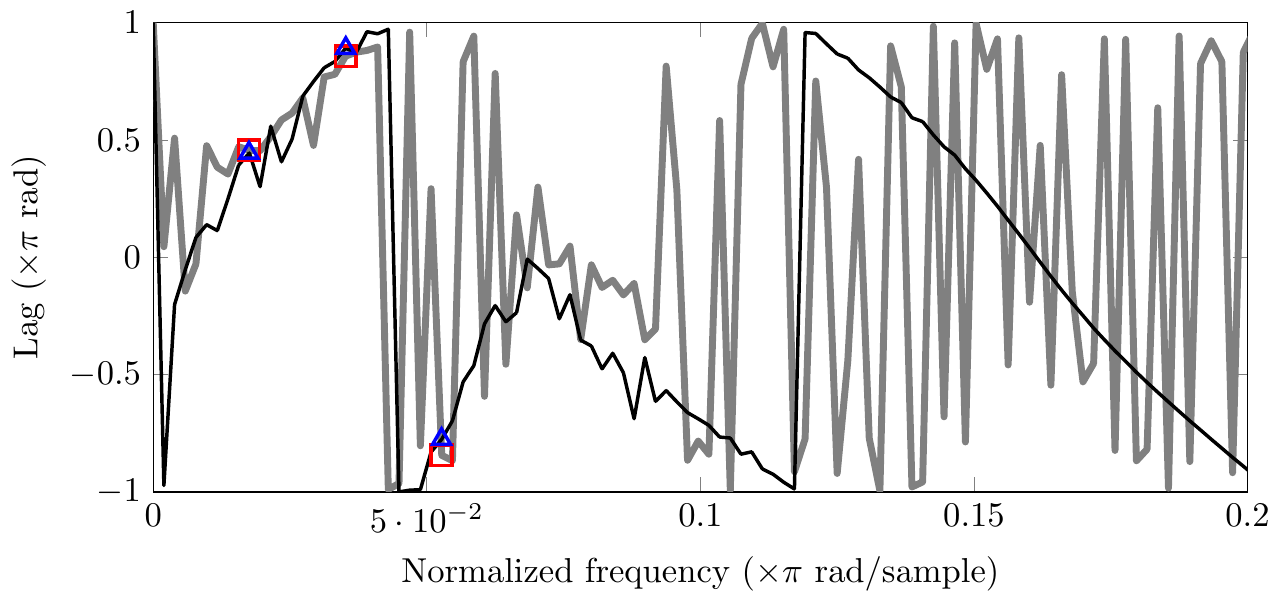}
\caption{
        CPSD of $p(\bm{x}_1,t)$ and $p(\bm{x}_2,t)$
        where $\bm{x}_1 = (2,-0.5)$
        and $\bm{x}_2 = (0,-0.5)$
        of channel driven cavity at ${\rm Re} \approx 5500$.
        DNS (thick gray line),
        stabilized $n,p=20$ ROM (black line)
        }\label{fig:high_Re_cavity_CPSD}
\end{figure}

\begin{figure}
\centering
    \includegraphics{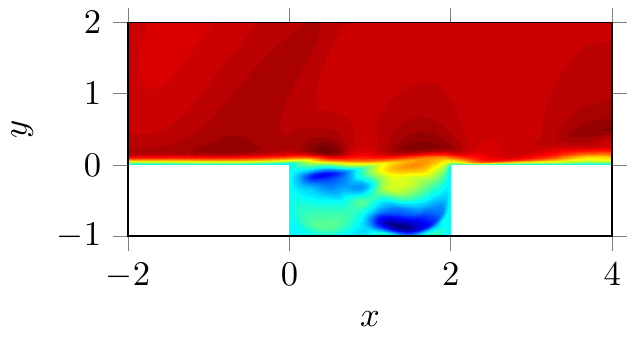}

    \includegraphics{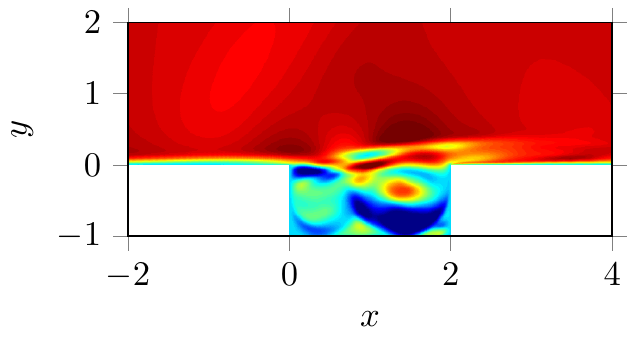}

    \includegraphics{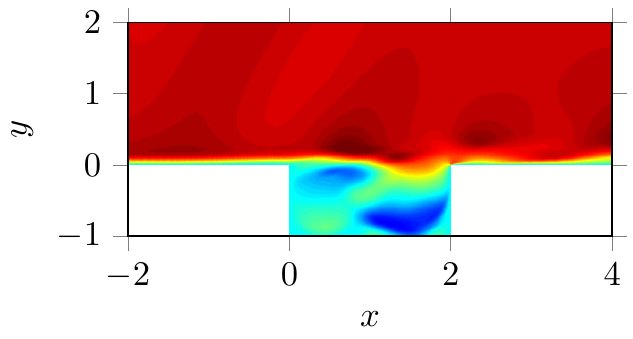}
\caption{
        Snapshot of channel drive cavity ${\rm Re} \approx 5500$;
        contours of $u$-velocity magnitude at the final snapshot.
        DNS (top),
        standard $n=20$ ROM (middle),
        and stabilized $n,p=20$ ROM (bottom)
        }\label{fig:high_Re_cavity_velocity_snapshot}
\end{figure}

\begin{figure}
\centering
    \includegraphics{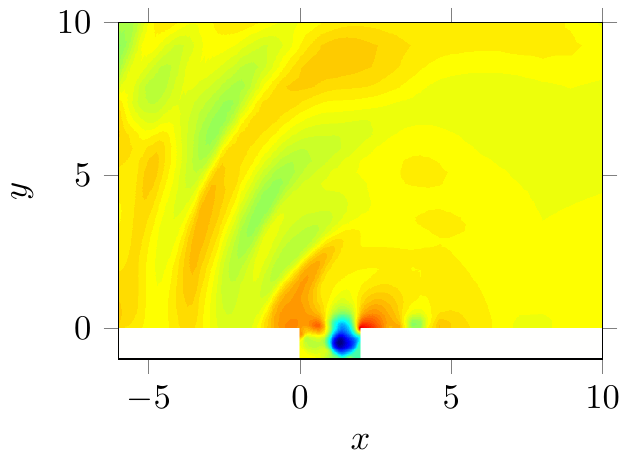}

    \includegraphics{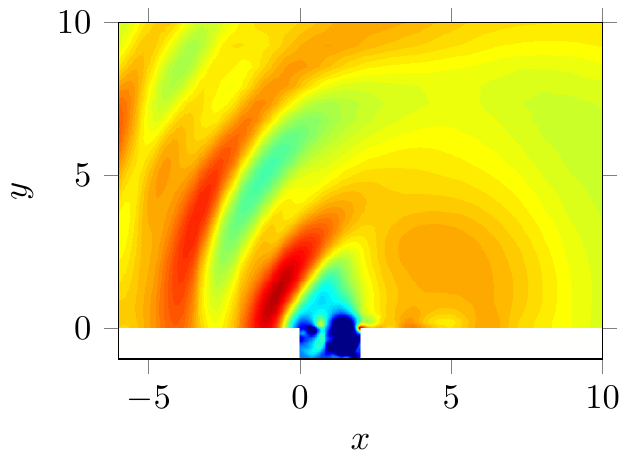}

    \includegraphics{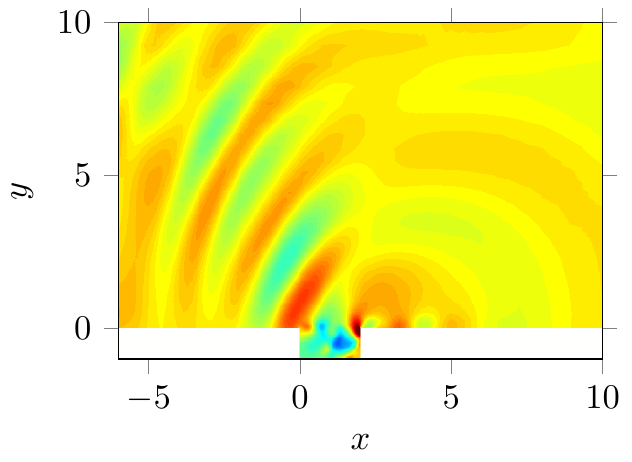}
\caption{
        Snapshot of channel driven cavity ${\rm Re} \approx 5500$;
        contours of pressure at the final snapshot.
        DNS (top),
        standard $n=20$ ROM (middle),
        and stabilized $n,p=20$ ROM (bottom)
        }\label{fig:high_Re_cavity_pressure_snapshot}
\end{figure}

Figures~\ref{fig:high_Re_cavity_PSD1} and~\ref{fig:high_Re_cavity_PSD2} shows
the PSDs of the predicted pressure fluctuations at locations $\bm{x}_1=(2,-0.5)$
and $\bm{x}_2=(2,0.5)$, respectively. The PSD of the CFD signal was computed
using all available snapshots from $t=0$ to $t=67$ where $t$ is non-dimensional.
On the other hand, the PSD of the stabilized ROM was computed from the signal
$10\times$ past the duration of the original snapshots; i.e. $t=(670-67)$ to
$t=670$.  The stabilized ROM accurately predicts the chaotic pressure
fluctuations at both locations.  Figure~\ref{fig:high_Re_cavity_CPSD}
illustrates the Cross Power Spectral Density (CPSD) for pressure fluctuations at
$\bm{x}_1$ and $\bm{x}_2$. Both the power and phase lag at the fundamental
frequency, and the first two super harmonics (normalized frequency ($\times \pi$
rad/sample) $\approx 0.18,0.35,$ and $0.53$) are predicted accurately using the
stabilized ROM. The phase lag at these three frequencies in
Figure~\ref{fig:high_Re_cavity_CPSD} as predicted by the CFD and the stabilized
ROM is identified by red squares and blue triangles, respectively. As expected,
the low-dimensional ROM is unable to reproduce the phase lag of low-amplitude
frequencies or higher-order super harmonics.  \clearpage Finally, a snapshot of
the predicted velocity and pressure magnitudes at the final snapshot are
illustrated in Figure~\ref{fig:high_Re_cavity_velocity_snapshot}
and~\ref{fig:high_Re_cavity_pressure_snapshot}. Since the flow at this higher
Reynolds number is chaotic, the low-dimensional model can not be expected to
track the original snapshots exactly.  However, the snapshots demonstrates that
the stabilized ROMs faithfully reproduce the large features of the flow. The
same cannot be said of the standard ROMs.

\subsection{Computational speed-up}

For each problem considered, the speed-up factor delivered by its ROM for the
online computations is reported in Table~\ref{tab:speed_up}.  All ROMs are
solved in MATLAB using the variable order solver \verb=ODE15s=; Jacobians are
provided analytically. For more details on this algorithm, the reader is
referred to \citet{Shampine_SIAM_JSC_1997}.  All online time-integration CPU
times were measured using the \verb=tic-toc= function on a single computational
thread via the \verb=-singleCompThread= start-up option. The FOM
time-integration, basis construction and Galerkin projection times given in the
table are reported in CPU-hours, calculated as the product of the number of
processors used in the computation and the mean CPU time over all processors.
The number of processors employed varied between 1 and 128. The online speed-up
is calculated by evaluating the ratio between the time-integration of the FOM
and the time-integration of the ROM. The reader can observe that the ROM online
speedup is on the order of at least $10^4$ for all three problems considered.
Moreover, the stabilization step takes very little time (on the order of
seconds/minutes).

\begin{table}
    \centering
\caption{CPU times for off-line and on-line computations.}
\begin{tabular}{l|c|c|c|}
\cline{2-4}
                                                        & \multicolumn{3}{c|}{Numerical Experiment}                                                                                                                                                         \\ \hline
\multicolumn{1}{|c|}{Procedure}                      &
\multicolumn{1}{c|}{Airfoil} &
\multicolumn{1}{c|}{\begin{tabular}[c]{@{}c@{}}Cavity,\\
Low-Re\end{tabular}} &
\multicolumn{1}{c|}{\begin{tabular}[c]{@{}c@{}}Cavity,\\
Moderate-Re\end{tabular}} \\ \hline \hline
\multicolumn{1}{|l|}{FOM \# of DOF}
    & 360,000 & 288,250 & 243,750  \\ \hline
\multicolumn{1}{|l|}{Time-integration of FOM}
    & 7.8 hrs & 72 hrs & 179 hrs  \\ \hline
\multicolumn{1}{|l|}{Basis construction (size $n+p$ ROM)}
    & 0.16 hrs & 0.88 hrs&  3.44 hrs        \\
\hline \multicolumn{1}{|l|}{Galerkin projection (size $n+p$ ROM)}
    & 0.74 hrs & 5.44 hrs& 14.8 hrs \\
\hline \multicolumn{1}{|l|}{Stabilization}
    & 28 sec & 14 sec & 170 sec \\
\hline \multicolumn{1}{|l|}{ROM \# of DOF}
    & 4 & 4 & 20  \\
\hline \multicolumn{1}{|l|}{Time-integration of ROM}
    & 0.31 sec & 0.16 sec & 0.83 sec \\
\hline \hline
\multicolumn{1}{|l|}{Online computational speed-up} &
$9.1\times10^4$              & $1.6 \times 10^6$          & $7.8
\times 10^5$
\\ \hline
\end{tabular}
\label{tab:speed_up}
\end{table}
\section{Conclusions}
\label{sect:concl}

In this paper, an approach for stabilizing and enhancing
projection-based fluid ROMs of the compressible Navier-Stokes
equations is developed.  Unlike traditional approaches, no empirical
turbulence modeling terms are required, and consistency between the
ROM and the full order model from which the ROM is derived is
maintained.  Mathematically, the approach is formulated as a
trace minimization on the Stiefel manifold. The method is
shown to yield both stable and accurate low-dimensional models of
several representative compressible flow problems. In particular,
the method is demonstrated on flows at higher Reynolds number where
the dynamics are chaotic. Future work will include the extension of
the proposed approach to problems with generic non-linearities,
where the ROM involves some form of hyper-reduction (e.g., DEIM,
gappy POD) following the procedure described in Remark 2,
as well as to predictive applications with varying Reynolds number
and geometry.


\clearpage
\appendix
\section{Construction of Galerkin matrices for the compressible Navier-Stokes
equations}\label{sec:Appendix_A}

In this section the construction of Galerkin matrices for the
compressible Navier-Stokes equations
\eqref{eqn:NS_eq} is outlined. Consider the standard
orthonormal POD vectors $u^i$, $v^i$, $\zeta^i$, and $p^i$,
$i=0,\cdots,n$ where $i=0$ identifies the constant mean flow. The following
products are generated for $j,k=0,\cdots,n$:

\begin{subequations}
    \begin{align}
    \phi_1^{(j,k)} &= u^j \odot \zeta_x^k + v^j \odot \zeta_y^k - u_x \odot \zeta^k   - v_y^j \odot \zeta^k, \\
    \phi_2^{(j,k)} &= u^j \odot u_x^k     + v^j \odot u_y^k     + \zeta^j \odot p_x^k - \frac{1}{{\rm Re}} \zeta^j \odot
        \left[
            \left(
                \frac{4}{3}u_x^k - \frac{2}{3}v_y^k
            \right)_x
            + (v_x^k + u_y^k)_y
        \right], \\
    \phi_3^{(j,k)} &= u^j \odot v_x^k + v^j \odot v_y^k + \zeta^j \odot p_y^k - \frac{1}{{\rm Re}} \zeta^j \odot
        \left[
            \left(
                \frac{4}{3}v_y^k - \frac{2}{3}u_x^k
            \right)_y
            + (v_x^k + u_y^k)_x
        \right], \\
    \begin{split}
    \phi_4^{(j,k)} &=  u^j \odot p_x^k + v^j \odot p_y^k + \gamma p^j \odot (v_x^k + u_y^k) -
        \frac{\gamma}{{\rm Re}\, {\rm Pr}}
        \left[
            (p^j \odot \zeta_k)_{xx} + (p^j \odot \zeta_k)_{yy}
        \right] \\
        &+ \frac{1-\gamma}{{\rm Re}}
        \left[
            u_x^j \odot
            \left(
                \frac{4}{3} u_x^k - \frac{2}{3} v_y^k
            \right)
            + v_y^j \odot
            \left(
                \frac{4}{3} v_y^k - \frac{2}{3} u_x^k
            \right)
            + (u_y^j +v_x^j) \odot (u_y^k +v_x^k)
        \right],
    \end{split}
    \end{align}
\end{subequations}

\noindent where subscripts identify partial spatial derivatives, and
$\odot$ is the Hadamard element-by-element product. A standard $L^2$ Galerkin
projection for $i,j,k=0,\cdots,n$ is performed as follows:

\begin{equation}
        b^{(i)}(j,k) = \sum_{m=0}^N
        \left[
        h \odot
        \left(
        \zeta^i \odot \phi_1^{(j,k)} +
        u^i \odot \phi_2^{(j,k)} +
        v^i \odot \phi_3^{(j,k)} +
        p^i \odot \phi_4^{(j,k)}
        \right)
        \right]_m,
\end{equation}
\noindent where $h$ is a vector of element volumes. Finally, the
standard Galerkin matrices for $i,j,k=1,\cdots,n$ with $a_0=1$, are given
by
\begin{subequations}
\begin{align}
    C_i            &= b^{(i)}(0,0), \\
    L_{i,j}        &= b^{(i)}(j,0) + b^{(i)}(0,j), \\
    Q^{(i)}_{j,k}  &= b^{(i)}(j,k).
\end{align}
\end{subequations}

\section*{Acknowledgments}

This material is based upon work supported by the National Science
Foundation under Grant No. NSF-CMMI-14-35474.  The authors would
like to thank Dr. Srinivasan Arunajatesan and Dr. Matthew Barone at
Sandia National Laboratories for some useful discussions on model
reduction for compressible flows.  The authors would also like to
thank Dr. Srinivasan Arunajatesan for generating the snapshots from
which the channel-driven cavity reduced order models examined in the
``Numerical experiments'' section of this paper were constructed,
and Dr. Jeffrey Fike for implementing the nonlinear
compressible Navier-Stokes equations in specific volume code in the
{\tt Spirit} code.



\end{document}